\documentclass[twocolumn,aps,prd,superscriptaddress,showpacs,nofootinbib]{revtex4-1}
\usepackage{epsfig,dcolumn,bm}

\newcommand{\dtres}{d^{\hspace{0.1mm} 3}\hspace{-0.5mm}}

\newcommand{\sqd}{\sqrt{2}}

\newcommand{\be}{\begin{eqnarray}}
\newcommand{\ee}{\end{eqnarray}}
\newcommand{\nn}{\nonumber}
\newcommand{\ddn}{D^0\bar D^{*0}}
\newcommand{\ddc}{D^+ D^{*-}}
\newcommand{\ddi}{D\bar D^{*}}
\newcommand{\kvec}{\vec{k}}
\newcommand{\pvec}{\vec{p}}
\newcommand{\kpvec}{\vec{k}^{\;\prime}}
\newcommand{\ppvec}{\vec{p}^{\;\prime}}
\newcommand{\xvec}{\vec{x}}

\newcommand{\Eq}[1]{Eq.~(\ref{#1})}

\newcommand{\psihat}{\hat \psi}

\begin{document}

\title{Couplings in coupled channels versus wave functions: application to the $X(3872)$ resonance}

\author{D. Gamermann}{\thanks{E-mail: daniel.gamermann@ific.uv.es}}

\author{J. Nieves}{\thanks{E-mail: jmnieves@ific.uv.es}}
\affiliation{Instituto de F\'isica corpuscular (IFIC), Centro Mixto
Universidad de Valencia-CSIC,\\ Institutos de Investigaci\'on de
Paterna, Aptdo. 22085, 46071, Valencia, Spain}

\author{E. Oset}{\thanks{E-mail: oset@ific.uv.es}}
\affiliation{Departamento de F\'isica Te\'orica and IFIC, Centro Mixto
Universidad de Valencia-CSIC,\\ Institutos de Investigaci\'on de
Paterna, Aptdo. 22085, 46071, Valencia, Spain}

\author{E. Ruiz Arriola}{\thanks{E-mail: earriola@ugr.es}}
\affiliation{Departamento de F\'isica At\'omica, Molecular y Nuclear,
Universidad de Granada, E-18071 Granada, Spain}

\begin{abstract}
We perform an analytical study of the scattering matrix and bound
states in problems with many physical coupled channels. We establish
the relationship of the couplings of the states to the different
channels, obtained from the residues of the scattering matrix at the
poles, with the wave functions for the different channels. The
couplings basically reflect the value of the wave functions around the
origin in coordinate space. In the concrete case of the $X(3872)$
resonance, understood as a bound state of $\ddn$ and $\ddc$ (and
$c.c.$\footnote{From now on, when we refer to $\ddn$, $\ddc$ or $\ddi$
we are actually referring to the combination of these states with
their complex conjugate in order to form a state with positive
C-parity.}), with the $\ddn$ loosely bound, we find that the couplings
to the two channels are essentially equal leading to a state of good
isospin $I=0$ character. This is in spite of having a probability for finding
the $\ddn$ state much larger than for $\ddc$ since the loosely bound
channel extends further in space. The analytical results, obtained
with exact solutions of the Schr\"odinger equation for the wave
functions, can be useful in general to interpret results found
numerically in the study of problems with unitary coupled channels
methods.
\end{abstract}


\maketitle


\section{Introduction}

The $X(3872)$ resonance, observed by Belle \cite{belle} and confirmed
by CDFII, D0 and BaBar collaborations \cite{cdf,d0,babar}, has been
the object of intense debate from the theoretical point of view (see
recent workshop on charm exotics at Badhonef
\cite{badhonef}). Although different tentative explanations to its
nature have been provided \cite{qq1,qq2,qq3,tornq,suzuki,closeypage,meuax} the idea most
supported recently is that it corresponds to a loosely bound state of
$\ddi$
\cite{meuax,mol1,mol2,mol3,mol4,mol5,mol6,mol7,mol8,mol9,mol10,mol11,mol12,
Matheus:2009vq,Lee:2009hy}
or slightly unbound, virtual $\ddi$ state
\cite{hanhart,meutwox}. However, the energy of the resonance is very
close to the $\ddn$ threshold, with the eventual charged components
$\ddc$ bound by about 8 MeV. The binding of the $\ddn$ could be so
small as to render the relatively very bound charged components
irrelevant, at least from the probability point of view, given the
fact that the loosely bound component would extend much further in
space than the charged components. This is the idea behind many works
\cite{mol4,mol6,mol8,mol9,mol10,mol12}. However it was found in
\cite{meutwox} that the couplings of the resonance to the charged and
neutral components were practically identical, implying a near $I=0$
nature of the resonance as experimentally established. A pure $\ddn$
component would have an equal admixture of $I=0$ and $I=1$ and, according
to \cite{meutwox} would produce a ratio of
\be
\frac{{\cal B}(X\rightarrow J/\psi\pi^+\pi^-)}
{{\cal B}(X\rightarrow J/\psi\pi^+\pi^-\pi^0)}\label{ratiopipi} \, , 
\ee
much larger than experiment. Indeed, in \cite{thomas} several works based on pion exchange
as the source for the $\ddi$ binding are analysed thoroughly, stressing the 
importance of taking into account the neutral and charged components 
to properly study isospin violation in the $X(3872)$.
There seems to be a contradiction between
the intuitive idea of a dominance of the loosely bound component and
the fact that experiment demands clearly an important contribution
from the charged components. The clarification of this puzzle and
establishing the meaning of the couplings in terms of wave functions
is the purpose of this present work. While in the theoretical
calculations one normally uses field theoretical methods to evaluate
observables \cite{meuax}, without resorting to wave functions, the
clarification of the puzzle forces one to face this problem solving
the Schr\"odinger equation for the wave functions in coupled channels.

For simplicity, we will assume that the $X(3872$) mass is below both
the $\ddc$ and $\ddn$ thresholds. The work proceeds as follows: in the
next section we make a brief summary of \cite{meutwox} to expose the
problem. In sections III and IV we solve the Schr\"odinger equation in
the case of one and two channels. In section V we extend the findings
to the case of many channels. In section VI we come back to the
$X(3872)$ and comment on its decay to $J/\psi$ plus two and three pions,
in section VII we comment on the independence of the results
with the choice of the potential and in section VIII we outline our conclusions.

\section{The $X(3872)$ within coupled channels $\ddn$ and $\ddc$}
\label{sec:qft}

In \cite{meuax} the $X(3872)$ was plausibly explained as an $I=0$
dynamically generated state in coupled channels with positive
C-parity. However, in that work the charged and neutral $D$ mesons were
put with the same mass, in which case one had a good isospin
symmetry. In \cite{meutwox} the masses were taken different and a
small isospin breaking was produced. Summarizing the approach of
\cite{meuax} we call channels 1 and 2 the $\ddn$ and $\ddc$. It was
found in this work, using the hidden gauge Lagrangians adapted to the
$SU(4)$ flavor symmetry, containing explicit breaking of the symmetry,
that the potential in coupled channels, in
s-wave, was very close to the type
\be
V^{\rm FT}&=&\left(\begin{tabular}{cc} 
$v^{\rm FT}$ & $v^{\rm FT}$ \\ $v^{\rm FT}$ &$v^{\rm FT}$ 
\end{tabular}\right)\label{pot},
\ee
the label ${\rm FT}$ standing for field theoretical approach, and
$v^{\rm FT}$ is, in principle, a function of the invariant mass $s$
(see Eqs.~(7) and (12) of Ref.~\cite{meutwox}). To describe the
dynamics of the $X(3872)$, which is placed quite close to the $\ddi$
threshold, it is sufficient to take the potential given in Eq.~(7) of
Ref.~\cite{meutwox}) at threshold, neglecting in the potential all isospin
breaking corrections induced by the difference of masses between
charged and neutral mesons\footnote{We will keep those isospin
breaking corrections in the loop function $G$ that will be introduced
below. As it was discussed at length in Ref.~\cite{meutwox}, they turn
out to be quite relevant.}
\be
v^{\rm FT} = - \frac{m_Dm_{D^*}}{f_D^2} \label{eq:pepe1}
\ee
with $f_D \sim 165$ MeV, the $D-$meson decay constant, and $m_D$ and
$m_{D^*}$ averages of the neutral and charged $D$ and $D^*$ meson
masses, respectively. The above interaction, should be considered with
an ultraviolet cutoff in momentum space of natural size for hadron
interactions, $\Lambda < 1$ GeV. In numerical calculations we use $
m_{D^0}=1865$ MeV, $ m_{\bar D^{*0}}=2007$ MeV, $ m_{D^+}=m_{D^-}=1870
$ MeV and $ m_{D^{*+}}=m_{D^{*-}} =2010$ MeV.

\begin{figure}
\begin{center}
\includegraphics[width=6cm,angle=-0]{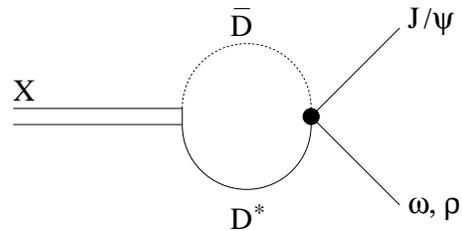}
\caption{$X$ decay mechanism into $J/\psi\omega(\rho)$ assuming the
$X$ to be a $\ddi$ molecule.} \label{fig1}
\end{center}
\end{figure}

\begin{figure}
\begin{center}
 \rotatebox{90}{\includegraphics[width=8cm,angle=-90]{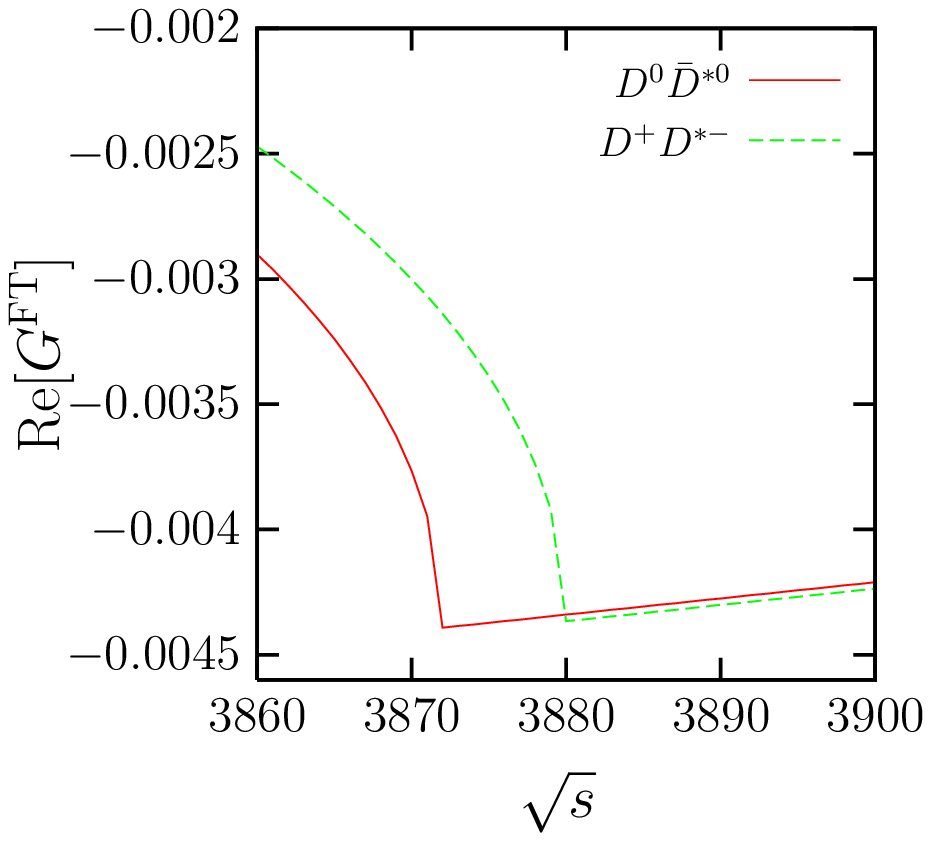}}
\caption{ Loop function $G^{\rm FT}$ for $\ddn$ and $\ddc$ from \cite{meutwox}.} \label{fig2}
\end{center}
\end{figure}

The Bethe-Salpeter equation
in coupled channels in the on-shell factorization approach stemming
from the use of the $N/D$ method \cite{noverd,meissner,Nieves:1999bx}
is given by:
\be
T^{\rm FT}&=&(1-V^{\rm FT}G^{\rm FT})^{-1}V^{\rm FT}\label{bseq}
\ee
where both $V^{\rm FT}$ and $T^{\rm FT}$ are on-shell\footnote{The
  normalization is fixed thanks to the relation between the scattering
  matrix and the differential center of mass cross section, 
\be
\frac{d\sigma}{d\Omega}\Big |_{\rm CM}= \frac{1}{64\pi^2
  s}\left|T^{\rm FT}\right|^2.
\ee
}, and $G^{\rm FT}$ is the diagonal loop function for the two
intermediate $D$ and $\bar D^*$ meson propagators. In the particular
case of Eq.~(\ref{pot}), Eq.~(\ref{bseq}) is trivially written as
\be
T^{\rm FT}&=&\frac{V^{\rm FT}}{1-v^{\rm FT}G^{\rm FT}_{11}-v^{\rm
    FT}G^{\rm FT}_{22}}
\ee
We will assume that the $T^{\rm FT}$-matrix develops a pole, in the
first Riemann sheet and below both thresholds, for the $X(3872)$
resonance and the couplings $g^{\rm FT}_i$ to the channels $\ddi$ are
defined such that in the vicinity of the pole
\be
T^{\rm FT}_{ij}&=&\frac{g^{\rm FT}_i g^{\rm FT}_j}{s-s_R}
\ee
with $s_R$ the squared mass of the resonance, which allows one to obtain
the couplings via
\be
g^{\rm FT}_i g^{\rm FT}_j&=& \lim_{s\rightarrow s_R} (s-s_R)T^{\rm
  FT}_{ij}=\left.\frac{-1}{ 
\frac{d}{ds}\left(G^{\rm FT}_{11}+G^{\rm
  FT}_{22}\right)}\right|_{s=s_R} \\
&=& [g^{\rm FT}]^2
\label{lhopt}
\ee
All the couplings are equal in this case and, as we see in
Eq.~(\ref{lhopt}), they are independent of $v^{\rm FT}$ and only
depend on the derivative of the loop function $G^{\rm FT}$. One is
tempted to interpret the couplings as the components of the wave
function, and since the couplings are equal we would have
\be 
|X(3872)\rangle &\propto& |\ddn\rangle  + |\ddc\rangle  
\ee 
which represents a
pure isospin $I=0$ state. Such an interpretation has some basis since the
equality of the couplings is what makes the state behave as an $I=0$
state in the field theoretical approach. Indeed, think of the ratio of
Eq.~(\ref{ratiopipi}). The two pion and the three pion states in the
decay of the $X(3872)$ correspond to a $\rho$ and an $\omega$
respectively, according to experiment \cite{cdf2,belle2}. In a field
theoretical approach the mechanism for this decay is depicted in 
Fig.~\ref{fig1}, and the ratio for $\rho$ and $\omega$ decays would be given
by\footnote{Technically, there can be a multiplicative coefficient in
  \Eq{rhoom}, but in the model used in \cite{meutwox} it is equal to
  1.}
\be
R_{\rho/\omega}&=&\left(\frac{g_1^{\rm FT} G^{\rm FT}_{11}-g^{\rm FT}_2 G^{\rm FT}_{22}}{g^{\rm FT}_1 G^{\rm FT}_{11}+g^{\rm FT}_2 G^{\rm FT}_{22}}\right)^2.\label{rhoom}
\ee
The plus and minus signs in the numerator and denominator of
Eq.~(\ref{rhoom}) are simply a consequence of the fact that
$J/\psi\rho$ has $I=1$ while $J/\psi\omega$ has $I=0$. If we had equal
masses for the charged and neutral $D$ mesons, the numerator of
Eq.~(\ref{rhoom}) vanishes and the decay $X\rightarrow J/\psi\rho$ is
forbidden, since it violates isospin. If the masses of the two
channels are different, even taking the two couplings equal, as we
found in Eq.~(\ref{lhopt}), the numerator of Eq.~(\ref{rhoom}) does
not vanish due to the difference in the loop function for each channel, see
Fig. \ref{fig2}, and the decay $X\rightarrow J/\psi\rho$ is
allowed. One could interpret this by saying that the $X(3872)$ is an
$I=0$ state but the intermediate loops in the decay violate isospin.
However, the free Hamiltonian (including masses) of the $\ddn$ and $\ddc$
system does not commute with isospin, which implies that the $X(3872)$ does not
have a well defined isospin. The decays of the $X(3872)$ resonance can
provide information on this isospin mixture. In this work we will look at
its $J/\psi\rho$ and $J/\psi\omega$ decays, assuming transition operators
of zero (short) range. Within this scheme, we will show that the isospin
violation in these decays is linked to the different probability amplitudes
of finding the $\ddn$ or $\ddc$ meson pairs, which form the $X$ molecule, at short
relative distances. This is the same as stating that the $\ddn$ and $\ddc$ wave functions
around the origin will determine the isospin violations for these decays.
Hence the couplings should not be understood as a measure of the wave function components.
We shall come back to this issue in what follows.

\section{Coupling and wave function in the one channel case}

We will first study the non-relativistic dynamics of a bound state
generated by the interaction of two particles of masses $m_1$ and
$m_2$, respectively.

\subsection{The Lippmann Schwinger equation}
\label{sec:qm}

We need a potential $V$ and to illustrate our results, which are
general, we can take some easy form for it (other forms will be analyzed in section \ref{sec:otherf}).
We choose a separable function in momentum
space with the modulating factor being a simple step function,
$\Theta$. Thus our potential, already projected in $s-$wave, is
assumed to be\footnote{We use normalization $\langle \vec p | \vec x
  \rangle = e^{-{\rm i} \vec p \cdot \vec x} /(2\pi)^{\frac32}$ which means
  $\int \dtres x | \vec x \rangle \langle \vec x | = \int \dtres p | \vec p
  \rangle \langle \vec p | = {\bf 1} $ so that for a {\it local}
  potential we have $\langle \ppvec|V|\vec{p}\,\rangle = \int \dtres x
  e^{-{\rm i} ( \ppvec -\vec p )\cdot \vec x} V(\vec x)/(2\pi)^3$.}
\be
\langle \ppvec|V|\vec{p}\,\rangle =& V(\ppvec,\vec{p}\,)=& v \,\Theta(\Lambda-p)\Theta(\Lambda-p^\prime)\label{potms}
\ee
where $p$ and $p^\prime$ stand for $|\vec{p}\,|$ and $|\ppvec|$ and
$\Lambda$ is a cutoff in momentum space.

Let the Hamiltonian be $H=H_0+V$ with $H_0$ the free Hamiltonian. In
this case the non-relativistic Lippmann Schwinger equation can be
written as

\be
T&=&V+V\frac{1}{E-H_0}T\label{lseq1}
\ee
or also as

\be
T&=&V+V\frac{1}{E-H}V\label{lseq2}
\ee

Taking Eq.~(\ref{lseq1}) we can write:
\be
\langle \vec{p}\,|T|\ppvec\rangle &=&\langle \vec{p}\,|V|\ppvec\rangle
+\nn\\
 &&\hspace{-0.5cm}\int_{k<\Lambda}\dtres k \,
 \frac{ \langle\, \vec{p}\,|V|\vec{k}\rangle}{E-m_1-m_2-\frac{{\vec k}^{\,2}}{2\mu}}\langle \vec{k}|T|\ppvec\rangle   \label{tmat}
\ee
where $\mu$ is the reduced mass of the two particles that interact 
[$1/\mu = 1/m_1+1/m_2$].

Eq.~(\ref{tmat}) has solution
\be
\langle \vec{p}\,|T|\ppvec\rangle
&=&\Theta(\Lambda-p)\Theta(\Lambda-p^\prime)\, t \label{eq:thetas}
\ee
which can also be seen from Eq.~(\ref{lseq2}), with $t$ given by

\be
t&=&v+v\,G\,t, \qquad  t = \frac{v}{1-vG} \label{tmat2}\\
G&=&\int_{p<\Lambda}\dtres p\frac{1}{E-m_1-m_2-\frac{\vec{p}^{\,2}}{2\mu}} \label{eq13}
\ee
We can see that Eq.~(\ref{tmat2}) is like the on-shell factorized
equation (no integral left) of Eq.~(\ref{bseq}), and $G$ is indeed, up
to a factor\footnote{The non-relativistic
  reduction of the scattering matrix, potential and two particle
  propagator loop function introduced in previous Sect.~\ref{sec:qft}
  are related to those defined here by 
\be
v^{FT} = 32 \pi^3 \mu \sqrt{s}~ v,  \qquad T^{FT} = 32 \pi^3 \mu
\sqrt{s} ~t  \label{eq:jna}
\ee
\be
G^{FT} = \frac{G}{32 \pi^3 \mu \sqrt{s}} \label{eq:jnb}
\ee 
with 
\be
 \quad G^{\textrm{FT}}&=&{\rm i}\int\frac{d^4q}{(2\pi)^4}\frac{1}{q^2-m_1^2+i\epsilon}\frac{1}{(P-q)^2-m_2^2+{\rm i}\epsilon}\label{eq:jn1}
\ee
and $P_\mu P^\mu=s$.}, the non-relativistic reduction of the loop function of
two particle propagators regularized with a cutoff $\Lambda$, as is
also usually done in the studies of hadron interactions with the
on-shell Bethe Salpeter equation \cite{osetreview}. The name Bethe
Salpeter equation was adopted in \cite{osetreview} because there,
relativistic meson propagators are used. For very weak bindings, it
is sufficient to use the Lippmann Schwinger equation, which
is what we do here.

The poles of Eq.~(\ref{tmat2}) occur for
\be
1-vG&=&0\textrm{ ,} \label{polecond}
\ee
which will occur for some value $E_\alpha < m_1+m_2$ where we have a bound
state. If the energy of the state is known, the above equation fixes the
cut-off $\Lambda$, or conversely, if the cut-off is fixed, one can predict the
energy of the state.

In the work of \cite{meuax} the theoretical approach is based on the underlying hidden gauge formalism \cite{hidden1,hidden2,hidden3}.
One assumes that it provides the potential $V$ within a scale of momenta which is of the order of
the cut off assumed here. Yet, one can assume certain uncertainties in $V$ which can be compensated
with changes in the range $\Lambda$ in order to obtain the proper binding, as we show below.
Indeed, using \Eq{polecond}, invariance under renormalization leads to
\be
\frac{d}{d\Lambda}\left ( vG \right)&=&0\\
\frac{1}{v}\frac{dv}{d\Lambda}=-\frac{1}{G}\frac{dG}{d\Lambda}
\ee
and hence for a fixed $E_\alpha$, the bare potential should have a
dependence on the cutoff to compensate that of the loop function $G$.

\subsection{The couplings of the state}

The coupling in this case is defined as $g$ such that in the vicinity
of the pole the scattering matrix behaves as

\be
t&=&\frac{g^2}{E-E_\alpha}
\ee
and hence
\be
g^2=\lim_{E\rightarrow E_\alpha} (E-E_\alpha)t= -\left(\frac{dG}{dE}\right)^{-1}_{E=E_\alpha}\label{eq16}
\ee
where \Eq{tmat2} has been used for $t$ and the l'H\^opital's rule has
been applied in the second equation.

The integral for the $G$ function defined in Eq.~(\ref{eq13}) can be
performed analytically and we obtain
\be
G(E_\alpha)&=&-8\mu\pi\left(\Lambda-\gamma \arctan\Big(\frac{\Lambda}{\gamma}\Big)\right) \label{eq31} \\
\gamma&=&\sqrt{2\mu E_B^\alpha}
\ee
where $0< E_B^\alpha = m_1+m_2-E_\alpha $ is the binding energy of the
state $\alpha$. The above equation allows us, using Eq.~(\ref{eq16}),
to write the coupling $g$ as
\be
g^2&=&\frac{\gamma}{8\pi\mu^2\left(\arctan\Big(\frac{\Lambda}{\gamma}\Big)-\frac{\gamma\Lambda}{\gamma^2+\Lambda^2}\right)}.
\label{eq:eqg}
\ee
We note that $g$ has a very smooth dependence on the cutoff
$\Lambda$, which can even be removed ($\Lambda \to \infty$). The coupling
is mostly determined by the binding energy (see \Eq{eq33} below), specially in the
limit in which the latter one is much smaller than the cutoff, and
hence  $g$ is, to great extent, renormalization scheme independent.

\subsection{The wave function}

The Schr\"odinger equation is given by:
\be
H|\psi\rangle &=&E|\psi\rangle 
\ee
where $\psi$ is an eigenfunction of $H$, the full Hamiltonian. We can write:
\be
(H_0+V)|\psi\rangle &=&E|\psi\rangle  \\
|\psi\rangle &=&\frac{1}{E-H_0}V|\psi\rangle 
\ee
which has the solution
\be \langle \vec{p}\,|\psi\rangle &=&\int \dtres k \int \dtres
k^\prime\langle \vec{p}\,|\frac{1}{E-H_0}|\kpvec\rangle \nn \\
&\times&\langle \kpvec|V|\kvec\rangle \langle \kvec|\psi\rangle \\
&=&v\,\frac{\Theta(\Lambda-p)}{E-m_1-m_2-\frac{\vec{p}^{\,2}}{2\mu}}\int_{k<\Lambda}
\dtres k \langle \vec{k}|\psi\rangle \label{wavep} 
\ee
which gives us the wave function. Integrating Eq.~(\ref{wavep}) over
$\dtres p$, we obtain 
\be
1-vG(E) = 0 
\ee
which is the condition to find the pole given in
Eq.~(\ref{polecond}). 

In Eq.~(\ref{wavep}), we determined the state
wave function up to a constant, $\int_{k<\Lambda}
\dtres k\langle \vec{k}|\psi\rangle $, which can be fixed from 
the normalization condition. Let  $E_\alpha < m_1+m_2$ be  the solution of the
above quantization equation, its wave function will satisfy
\be
\int \dtres p\, |\langle \vec{p}\,|\psi\rangle |^2=1
\ee
Note, that the wave function can be normalized because we are dealing
with a bound state whose energy is below $m_1+m_2$. From the above
equation, one easily finds
\be
1&=&v^2\int_{p<\Lambda} \dtres p\left(\frac{1}{E_\alpha-m_1-m_2-\frac{\vec{p}^{\,2}}{2\mu}}\right)^2\nn \\
&\times&\left|\int_{k<\Lambda}\dtres k\langle \vec{k}|\psi\rangle \right|^2
\ee
and hence, it follows
\be
 \left|v\int_{k<\Lambda}\dtres k \langle \vec{k}|\psi\rangle \right|^2=-\left(\frac{dG}{dE}\right)^{-1}_{E=E_\alpha}\textrm{.} \label{eq22}
\ee
We can now use the form of Eq.~(\ref{lseq2}) to solve the $T$ matrix. We
would have
\be
T&=&V+\sum_{m,m^\prime}V|m\rangle \langle m|\frac{1}{E-H}|m^\prime\rangle \langle m^\prime|V \label{eq23}
\ee
where $|m\rangle $ and $|m^\prime\rangle $ are complete sets of
eigenstates of $H$. In the vicinity of the pole at $E=E_\alpha$ we
care only  for the contribution of channel $\alpha$,
\be \hspace{-0.5cm}\langle \pvec\,|T|\ppvec\rangle &\sim&\langle
\pvec\,|V|\alpha\rangle \frac{1}{E-E_\alpha}\langle
\alpha|V|\ppvec\rangle \nn \\ &=&\int \dtres k\int \dtres k^\prime
\langle \pvec\,|V|\kvec\,\rangle \langle \kvec|\alpha\rangle \nn\\
&\times&\frac{1}{E-E_\alpha}\langle \alpha|\kpvec\rangle \langle
\kpvec|V|\ppvec\rangle \nn\\
&=&\frac{\left|v\,\int_{k<\Lambda}\dtres k \langle
\vec{k}|\alpha\rangle \right|^2}{E-E_\alpha} \Theta(\Lambda-p)\Theta(\Lambda-p^\prime)\label{eq24} 
\ee
exhibiting the form of \Eq{eq:thetas} from where we defined $t$. 
We can obtain the residue of $t$ as
\be
g^2&=&\lim_{E\rightarrow
  E_\alpha}(E-E_\alpha)T=\left|v\,\int_{k<\Lambda}
\dtres k \langle
\vec{k}|\alpha\rangle \right|^2\nn\\
&=&-\left(\frac{dG}{dE}\right)^{-1}_{E=E_\alpha}  \label{eq25}
\ee
where we have used Eq.~(\ref{eq22}) to get the last equality ($\alpha$ stands for $\psi$ here) and we note that this is the same result as in Eq.~(\ref{eq16}).

The wave function in coordinate space can be equally evaluated:

\be
\langle \vec{x}|\psi\rangle &=&\int \dtres p \langle \vec{x}|\vec{p}\,\rangle \langle \vec{p}\,|\psi\rangle  \nn \\
&=& \int \frac{\dtres p}{(2\pi)^{3/2}} e^{i\vec{p}.\vec{x}} \langle \vec{p}\,|\psi\rangle . \label{eq26}
\ee
Using Eq.~(\ref{wavep}) we find\footnote{It is possible 
to write \Eq{eq27} in terms of the analytical
  functions sine integral and cosine integral.}

\be
\hspace{-0.2cm}\langle \vec{x}|\psi\rangle &=& g 
\sqrt{\frac{2}{\pi}} \frac{1}{r}{\rm Im}  
\int_0^\Lambda dp\, p \frac{e^{{\rm i}p
    r}}{E_\alpha-m_1-m_2-\frac{\vec{p}^{\,2}}{2\mu}} \label{eq27}
\ee
For large values of $r$ this function goes as
\be
\langle \xvec|\psi\rangle _{_{r\rightarrow\infty}}&\sim&\frac{A}{\sqrt{4 \pi} r}e^{-\gamma r} \label{eq28}
\ee
where 
\be 
\gamma &=& \sqrt{2\mu E_B^\alpha} \, , \\ 
A &=& - 2\mu g \sqrt{2} \pi\left \{ 1 + {\cal O}(1/\Lambda)\right\} \, . 
\label{eq:A-gamma} 
\ee The exponential fall-off at large distances is controlled by the
binding energy of the state $\alpha$ and the coupling of the
state. This behavior follows the general rule of bound states outside
the interaction region and, as we see, is largely independent of the
cut-off $\Lambda$. For the same reason, $A$ carries information on 
the interaction region.

\subsection{The meaning of the coupling in terms of wave functions}

By means of Eq.~(\ref{eq25}) we know that (assuming a real
wave-function for the bound state) 

\be g=v\int_{k<\Lambda}\dtres k \langle \vec{k}|\alpha\rangle , 
\ee
and from Eq.~(\ref{eq26}) we can obtain the value of the wave function
at the origin in coordinate space:
\be \langle \vec{x}=\vec{0}\,|\psi\rangle \equiv
\psi(\vec{0}\,)&=&\int \frac{\dtres p}{(2\pi)^{3/2}}\langle
\vec{p}\,|\psi\rangle \label{funcori} 
\ee 
and so
\be
g&=&(2\pi)^{3/2} G^{-1}(E_\alpha) \psi(\vec{0}\,) \label{eq29}
\ee
where we have also used the condition for the bound state $1-vG=0$.
Since $g$ hardly depends on the cutoff $\Lambda$, the wave function at
the origin inherits the linear dependence on $\Lambda$ exhibited by
$G(E_\alpha)$ in \Eq{eq31}. Yet, if $v$ is known, the binding energy
also fixes $G(E_\alpha)$ from the condition $vG(E_\alpha)=1$. 

Now we define
\be \hat{\psi}&=&gG(E_\alpha)=(2\pi)^{3/2}
\psi(\vec{0}\,). \label{eq30} 
\ee 
This constant will appear often in
what follows. This is an important result concerning the problem at
stake: the coupling, up to the factor $G(E_\alpha) $, is a measure of
the wave function in coordinate space at the origin.

\subsection{The limit of small bindings}

Taking the limit for small values of $\gamma$ in \Eq{eq:eqg} we see that
\be
\lim_{\gamma\rightarrow0}g^2&=&\frac{\gamma}{4\pi^2\mu^2} \label{eq33}\ee
a result well known (up to a normalization depending on definitions)
\cite{weinberg,weinberg2,baru}.

This result can also be obtained from the general form of the
scattering amplitude at low energies. Indeed, let us recall
the form of the for s-wave scattering amplitude, $f$, close but above
threshold,
\be 
\hspace{-0.3cm}f^{-1}(E)&=&k \cot\delta-{\rm i}\,k  \sim -\frac{1}{a}+\frac{r_0
  k^2}{2}+ \cdots -{\rm i}\,k,
\ee
with $\delta$ the phase shifts, $a$ and $r_0$ effective range
parameters and $k=\sqrt{2\mu(E-m_1-m_2)}$. The analytic continuation 
below threshold to the energy of the bound state
$E=E_\alpha$ reads (we assume $E_\alpha$ is close to threshold)
\be
f^{-1}(E_\alpha)=-\frac{1}{a}-\frac{r_0 \gamma^2}{2}+ \cdots\gamma \label{eq34}
\ee 
where we have taken $k(E_\alpha)=i\gamma$. The inverse of the
scattering matrix must vanish at $E=E_\alpha$,  as it corresponds
to a bound state, and  the limit:
\be
\hat{g}^2&=&\lim_{E\rightarrow E_\alpha}(E-E_\alpha)f(E)=
\lim_{E\rightarrow E_\alpha}\frac{E-E_\alpha}{f^{-1}(E)}\nn\\
 &=&\frac{-1}{\frac{d}{d
    E}\left(\frac{1}{a}+\frac{r_0\gamma^2}{2}-\gamma\right)_{E=E_\alpha}
} \sim-\frac{\gamma}{\mu} + {\cal O}(\gamma^2)\label{eq35} 
\ee 
which agrees with Eq.~(50) of Ref.~\cite{jnieves}. 

This result is equivalent to the one in Eq.~(\ref{eq33}) up to a
normalization which is easy to get recalling that 
\be 
f = - \frac{T^{FT}}{8\pi\sqrt{s}} = -4\pi^2\mu\,t
\ee
As a consequence the coupling that we are using becomes
\be
g^2=-\frac{\hat{g}^2}{4\pi^2\mu}=\frac{\gamma}{4\pi^2\mu^2} 
+ {\cal O}(\gamma^2)
\ee 
which is the result obtained in Eq.~(\ref{eq33}).

A final remark concerns the comparison of this result with the
couplings defined in \cite{meutwox} and in general in studies using
the chiral unitary approach \cite{osetreview} where $G$ is defined in
a field theoretical approach in terms of two relativistic propagators
(Eq.~(\ref{eq:jn1})). Note that from Eqs.~(\ref{eq:jna}) and 
(\ref{eq:jnb})
\be
G^{\rm FT}v^{\rm FT} = Gv \label{eq:jn2}
\ee
which guaranties that the position of the pole remains
unchanged, since it is determined by the condition $Gv=1$. Besides, from Eq.~(\ref{eq:jna}), we trivially find
\be
g^{\rm FT} &=& \left (64\pi^3 \mu E_\alpha^2 \right)^\frac12\, g \label{eq:g}\\
&\sim& 
E_\alpha\left (16\pi \gamma /\mu \right)^\frac12 \quad (\gamma\rightarrow 0) 
\ee
\section{Two coupled channels}
\label{sec:2-cc}

\subsection{The couplings}

We work out in this section the two channel problem for the particular
case of the $X(3872)$ using a dynamics determined by the potential of
Eq.~(\ref{pot}), which is an isoscalar operator (i.e., it is diagonal
in the isospin basis). We will work first
within the Quantum Mechanics formalism that is adequate here, since
the mass of the $X(3872)$ resonance and that of the charged and
neutral $\ddi$ pairs differs in just few MeV. We will use again a
cut-off $\Lambda$ in momentum space, and the $2\times 2$ matrices
$T$ and $V$ will encode step functions
\be
\langle \ppvec|V|\vec{p}\,\rangle \equiv& V(\ppvec,\vec{p}\,)=&
v \,\Theta(\Lambda-p)\Theta(\Lambda-p^\prime) \nn \\
\langle \vec{p}\,|T|\ppvec\rangle\equiv
&T(\ppvec,\vec{p}\,)=& t \, \Theta(\Lambda-p)\Theta(\Lambda-p^\prime) \label{potte}
\ee
with
\be 
v&=&\left(\begin{tabular}{cc} ${\hat v}$ & ${\hat v}$
  \\ ${\hat v}$ &${\hat v}$ \end{tabular}\right) \label{eq:vqm}
\ee
The Lippmann Schwinger equation in the coupled channel space reads
\be t&=&(1-vG)^{-1}v \label{eq:eq57bis}
\\ &=&\frac{1}{1-{\hat v}G_{11}-{\hat v}G_{22}}v  \label{eq:eq57}
\ee
where 
\be 
\hspace{-0.3cm}G &=&\left(\begin{tabular}{cc} $G_{11}$ & 0
  \\ 0 & $G_{22}$ \end{tabular}\right), \quad
G_{ii}=\int_{p<\Lambda}
\frac{\dtres p}{E-M_i-\frac{\vec{p}^{\,2}}{2\mu_i}}\label{eq39}
\ee
with $E$ the relative energy including the mass of the particles and
$M_1$ and $M_2$ the thresholds of each channel 
\be
M_1=m_{D^0}+m_{\bar
D^{*0}}, \qquad  M_2=m_{D^+}+m_{D^{*-}}
\ee
and $\mu_1$ and $\mu_2$ the reduced masses of the $\ddn$ and $\ddc$
systems respectively\footnote{\label{foot:avg}The correspondence
of the  Quantum Field Theory and Quantum Mechanics, in
the nonrelativistic limit, scattering matrix, two
particle propagator and potential matrices reads
\be 
G^{FT} &=& \frac{1}{32\pi^3\sqrt{s}}\, \mu^{-\frac12}\, G \,\mu^{-\frac12}\\
T^{FT} &=& 32\pi^3\sqrt{s}\,\mu^\frac12 \,t\, \mu^\frac12, \\
 V^{FT}& =& 32\pi^3\sqrt{s}\,\mu^\frac12 \,v
    \, \mu^\frac12 \label{eq:vft}
\ee
where 
\be 
\mu^\frac12  &=&\left(\begin{tabular}{cc} 
$\sqrt{\mu_1}$ & 0 \\ 0 &$\sqrt{\mu_2}$ 
\end{tabular}\right) \label{redmass}
\ee and $\mu^{-\frac12}$ the inverse of the above matrix. The Bethe
Salpeter equation~(\ref{bseq}) implies, in the non-relativistic
limit, the Lippmann Schwinger equation ~(\ref{eq:eq57bis}). Thus, it
trivially follows
\be
\det(1-V^{FT}\,G^{FT})=\det(1-v\,G)
\ee
which guaranties that poles are placed in the same position in both
approaches. Finally, note that a potential in the field theory
approach of the form in Eq.~(\ref{pot}) does not lead to a
quantum mechanics potential of the form assumed in Eq.~(\ref{eq:vqm}),
unless that both reduced masses $\mu_1$ and $\mu_2$ are taken to be equal
in Eq.~(\ref{redmass}). This is an excellent approximation for the case
of the coupled channels $\ddn$ and $\ddc$, and we have done so here when
relating Quantum Field Theory and Quantum Mechanics quantities. Thus,
we have taken $\mu_1\sim \mu_2 \sim \bar\mu$, with $\bar\mu$ some
average reduce mass. Hence, the Quantum Mechanics potential, ${\hat v}$, near
threshold and for an ultraviolet  cutoff of natural size for hadron
interactions $\Lambda < 1$ GeV, can be now approximated by  
\be
{\hat v} = -\frac{1}{32\pi^3f^2_D} \label{eq:vmq}
\ee
as deduced from \Eq{eq:pepe1}, with the further approximation
$\bar\mu\sqrt{s}\approx m_Dm_{D^*}$.}.
Assuming that the $X(3872)$ is a bound state of
the system, its mass ($E_\alpha \le M_1 < M_2$) will be obtained by
requiring that the denominator of Eq.~(\ref{eq:eq57}) will vanish
(pole of the $t-$matrix in the first Riemann sheet and below all
thresholds). Clearly when $E_\alpha\rightarrow M_1$ the $\ddn$ is
loosely bound and the $\ddc$ is bound by about 8 MeV. The explicit
expressions for $G_{ii}(E_\alpha)$ are given by Eq.~(\ref{eq31}) with
\be 
\gamma_i &=&\sqrt{2\mu_i E_{Bi}^\alpha} \\ 
E_{Bi}^\alpha &=& M_i-E_\alpha 
\label{eq:gamma-th}.
\ee 

Let us now pay attention to the couplings of the bound state to each
of the two channels.  Since all elements of the matrix $v$
are equal, both couplings $g_1$ and $g_2$ are the same:
\be g_1^2=g_2^2 \equiv  g^2&=&\lim_{E\rightarrow E_\alpha}(E-E_\alpha)t_{ij}
\nn\\ 
&=&-\left.
\left(\frac{dG_{11}}{dE}+\frac{dG_{22}}{dE}\right)^{-1}\right|_{E=E_\alpha}\label{eqres}
\ee
On the other hand, as in Eq.~(\ref{eq:g}), we obtain
\be
g^{\rm FT} &=& \left (64\pi^3 \bar\mu E_\alpha^2 \right)^\frac12\, g \label{eqres1}
\ee
with $\bar\mu$, the average of the $\mu_1$ and $\mu_2$ reduced masses,
as discussed in the footnote \ref{foot:avg}. By using
Eqs.~(\ref{eqres}) and (\ref{eqres1}),  we find not only a
qualitative, but also a quantitative good agreement  with the results
for the coupling shown in Fig. 3 of Ref.~\cite{meutwox}. For instance,
if the neutral channel is bound by 1 (0.1) MeV, the denominator of
Eq.~(\ref{eq:eq57}) will vanish for a value of the cutoff $\Lambda$
of around 680 (653) MeV, with ${\hat v}$ given by \Eq{eq:vmq}. This leads to
a coupling $g^{\rm FT }$ of the order of 5400 (3200) MeV, in reasonable
agreement with the result in Ref.~\cite{meutwox}.

In the limit when $E_{B1}^\alpha\rightarrow 0$, we have
$\left. \frac{dG_{11}}{dE}\right |_{E=E_\alpha}\rightarrow \infty$ and we find
\be
g_1^2=g_2^2&\sim &\frac{\gamma_1}{4\pi^2\mu_1^2}, \qquad
E_{B1}^\alpha\rightarrow 0   \label{eq40}
\ee
thus, both couplings go to zero as $\sqrt{\gamma_1}$.

\subsection{The wave function}

For the bound state we have now a two component wave function,
representing each of the $\ddn$ and $\ddc$ channels, and the
Schr\"odinger equation reads
\be
(H_0+V)|\psi\rangle &=&E|\psi\rangle  \label{eq41} \\
|\psi\rangle &=&\left(\begin{array}{c} |\psi_1\rangle  \\ |\psi_2\rangle  \end{array} \right)
\ee
The solution to this equation is given by
\be
|\psi\rangle &=&\frac{1}{E-H_0}V|\psi\rangle  \label{eq42} \\
\langle \pvec\,|\psi\rangle &=&\left(\begin{array}{cc}
  \frac{1}{E-M_1-\vec{p}^{\,2}/2\mu_1} & 0 \\ \\
                                                  0 & \frac{1}{E-M_2-\vec{p}^{\,2}/2\mu_2} \end{array}\right)\nn\\
&\times&\int \dtres k \langle \pvec\,|V|\kvec\rangle \langle \kvec|\psi\rangle  \label{eq43}
\ee
which represents two coupled channels equations
\be
\langle \pvec\,|\psi_1\rangle &=&{\hat v}\frac{\Theta(\Lambda-p)}{E-M_1-\frac{\vec{p}^{\,2}}{2\mu_1}}\nn\\
&\times&\int_{k<\Lambda}\dtres k\, \left(\langle \kvec|\psi_1\rangle
+\langle \kvec|\psi_2\rangle \right) 
\label{eq44a} \\
\langle \pvec\,|\psi_2\rangle &=&{\hat v}\frac{\Theta(\Lambda-p)}{E-M_2-\frac{\vec{p}^{\,2}}{2\mu_2}}\nn\\
&\times&\int_{k<\Lambda}\dtres k \, \left (\langle \kvec|\psi_1\rangle +\langle \kvec|\psi_2\rangle \right) \label{eq44b}
\ee
which require to know $\int_{k<\Lambda}\dtres k\langle
\kvec|\psi_i\rangle $ for its solution. To evaluate this latter
magnitude let us integrate over $\pvec$ in Eq.~(\ref{eq43}) and we get
\be
\int_{p<\Lambda}\dtres p\langle \pvec\,|\psi\rangle &=&G\,v\, \int_{p<\Lambda}\dtres p\langle \pvec\,|\psi\rangle  \label{eq45}
\ee
an algebraic equation that requires for its solution
\be
\det(1-G\,v)=&1-{\hat v}G_{11}-{\hat v}G_{22}=&0\label{eq46}
\ee
This equation is satisfied for the poles, $E=E_\alpha$, of the $t$ matrix corresponding
to bound states (see Eq.~(\ref{eq:eq57})). Eq.~(\ref{eq45}) can be now be
solved and we obtain 
\be
\int_{p<\Lambda} \dtres p\langle \pvec\,|\psi_2\rangle
&=&\frac{{\hat v}\,G^\alpha_{22}}{1-{\hat v}\,G^\alpha_{22}}\int_{p<\Lambda} \dtres p\langle \pvec\,|\psi_1\rangle  \nn\\
&=&\frac{G^\alpha_{22}}{G^\alpha_{11}}\int_{p<\Lambda} \dtres p\langle \pvec\,|\psi_1\rangle  \label{eq47}
\ee
where $G_{ii}^\alpha=G_{ii}(E=E_\alpha)$. Thus,  Eqs.~(\ref{eq44a})
and (\ref{eq44b}) can be written as
\be \langle \pvec\,|\psi_1\rangle
=\frac{1}{G^\alpha_{11}}\frac{\Theta(\Lambda-p)}{E_\alpha-M_1-\frac{\vec{p}^{\,2}}{2\mu_1}}\int_{k<\Lambda}\dtres
k\langle \kvec|\psi_1\rangle \label{eq48a} \\ \langle
\pvec\,|\psi_2\rangle
=\frac{1}{G^\alpha_{11}}\frac{\Theta(\Lambda-p)}{E_\alpha-M_2-\frac{\vec{p}^{\,2}}{2\mu_2}}\int_{k<\Lambda}\dtres
k\langle \kvec|\psi_1\rangle \label{eq48b} \ee
If we define the partial probability 
\be 
P_i = \langle \psi_i | \psi_i \rangle = \int \dtres p |\langle \pvec\,|\psi_i\rangle |^2
\ee
we can further use the total normalization condition
\be
1&=& P_1 + P_2 \equiv \int_{p<\Lambda} \dtres p\,\left\{|\langle \pvec\,|\psi_1\rangle |^2+|\langle \pvec\,|\psi_2\rangle |^2\right\} \nn\\
&=&-\left(\left.\frac{1}{[G_{11}^\alpha]^2}\frac{dG_{11}}{dE}\right|_{E=E_\alpha}+\left.\frac{1}{[G_{11}^\alpha]^2}\frac{dG_{22}}{dE}\right|_{E=E_\alpha}\right)\nn\\
&\times&\left|\int_{k<\Lambda} \dtres k\langle \kvec|\psi_1\rangle \right|^2 \label{eq49}
\ee
from where, using Eq.~(\ref{eqres})
\be
\left|\int_{p<\Lambda} \dtres p\langle \pvec\,|\psi_1\rangle \right|^2&=&[G_{11}^\alpha]^2\,g^2 \label{eq50}
\ee
and hence
\be
P_1 &=& -g^2 \left.\frac{dG_{11}}{dE}\right|_{E=E_\alpha},
\quad P_2 = 
-g^2 \left.\frac{dG_{22}}{dE}\right|_{E=E_\alpha} \\
\frac{P_1}{P_2} &=&
\frac{\mu_1^2\gamma_2}{\mu_2^2\gamma_1}\frac{\arctan\left(\frac{\Lambda}{\gamma_1
  }\right)-\frac{\gamma_1\Lambda}{\gamma_1^2+\Lambda^2}}{\arctan\left(\frac{\Lambda}{\gamma_2}\right)
-\frac{\gamma_2 \Lambda}{\gamma_2^2+\Lambda^2})}\nn\\
&=&\frac{\gamma_2}{\gamma_1} \left[ 1 + {\cal O} (\Lambda^{-1})  \right] 
= 
\sqrt{\frac{E_{B2}^\alpha}{E_{B1}^\alpha}} \left[ 1 + {\cal O} (\Lambda^{-1})  \right] 
\label{eq:ratio}
\ee
The neglected terms are finite range corrections, which in our case
are represented by the finite cut-off. On the other hand, assuming
real wave functions, Eq.~(\ref{eq50}) together with Eq.~(\ref{eq47})
lead to
\be
g\,G_{11}^\alpha=\int_{p<\Lambda} \dtres p\langle \pvec\,|\psi_1\rangle  \label{eq51a}\\
g\,G_{22}^\alpha=\int_{p<\Lambda} \dtres p\langle \pvec\,|\psi_2\rangle  \label{eq51b}
\ee
The wave functions in coordinate space would be again given by means
of Eq.~(\ref{eq27}) using $\mu_1$ for $\psi_1$ and $\mu_2$ for
$\psi_2$ and substituting $m_1+m_2$ by $M_i$ for each component. At
long distances this means, see Eq.~(\ref{eq28}),  
\be \langle \xvec|\psi_1 \rangle
_{_{r\rightarrow\infty}}&\sim&\frac{A_1}{\sqrt{4 \pi} r}e^{-\gamma_1
  r} \label{eq28'} \\ \langle \xvec|\psi_2 \rangle
_{_{r\rightarrow\infty}}&\sim&\frac{A_2}{\sqrt{4 \pi} r}e^{-\gamma_2
  r} \label{eq28''} \ee 
where we have used Eq.~(\ref{eq:gamma-th}).  Once again
$\int_{p<\Lambda} \dtres p\langle \pvec\,|\psi_i\rangle $ can be
interpreted (up to a constant factor $(2\pi)^\frac32$) as the wave
function at the origin, as done in Eq.~(\ref{funcori}) and then
\Eq{eq51a} and \Eq{eq51b} can be rewritten as
\be
gG^\alpha_{11}=&(2\pi)^{3/2}\psi_1(\vec{0}\,)=&\hat{\psi_1} \label{eq53a} \\
gG^\alpha_{22}=&(2\pi)^{3/2}\psi_2(\vec{0}\,)=&\hat{\psi_2} \label{eq53b} 
\ee
>From the above expressions, we trivially find the ratio of wave
functions at the origin
\be
\frac{\hat{\psi_2}}{\hat{\psi_1}} =
\frac{G^\alpha_{22}}{G^\alpha_{11}} = ({\hat v}G^\alpha_{11})^{-1}-1
\ee
and their departure from unity provides a tangible measure of the isospin
breaking in the interaction region.
On the other hand, 
\be
\frac{d}{d\Lambda} \left (\hat{\psi_2}/\hat{\psi_1}\right) =
\frac{\pi}{2}\frac{\mu_2}{\mu_1}\frac{\gamma_2-\gamma_1}{\Lambda^2} +
     {\cal O}(1/\Lambda^4) \label{eq:ratio-l}
\ee
which shows that, though both $\hat{\psi_1}$ and $\hat{\psi_2}$ depend
greatly on the cutoff, such a dependence is much reduced in their ratio.

\begin{figure}
\begin{center}
\begin{tabular}{c}
\includegraphics[width=7.5cm,angle=-0]{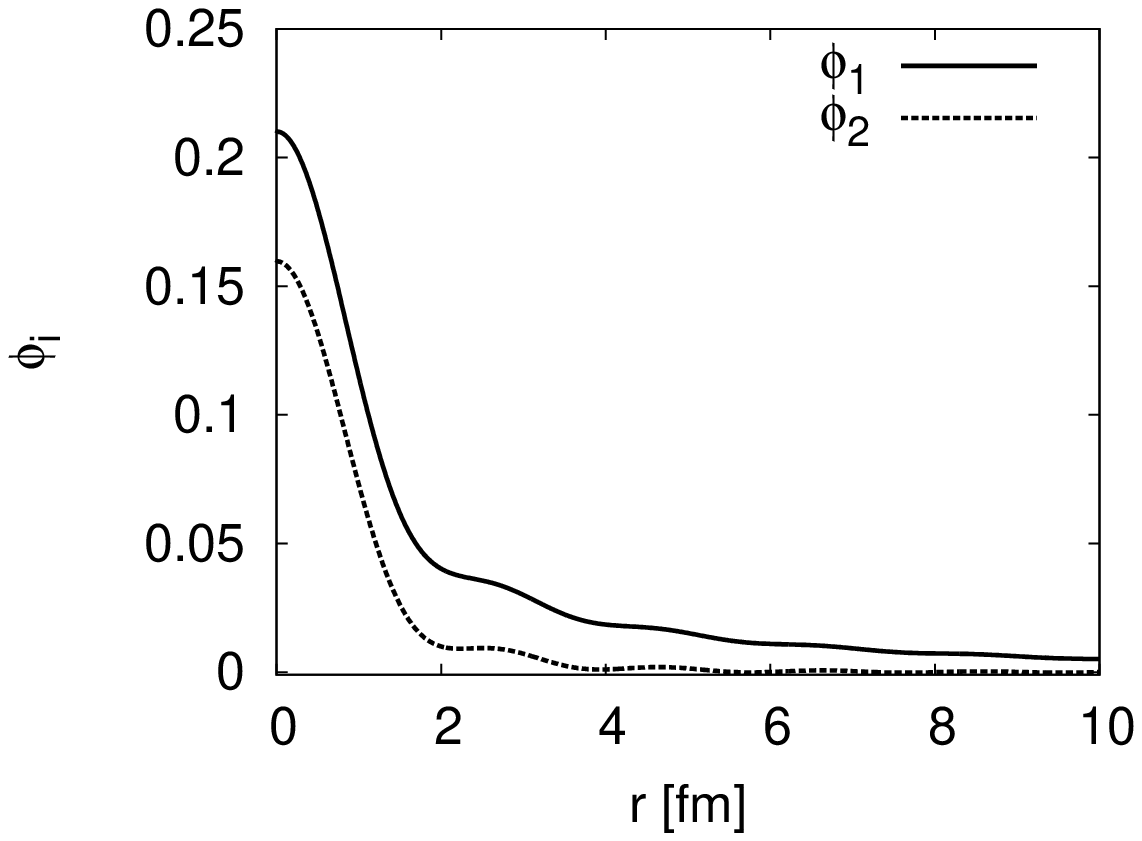} \\
\includegraphics[width=7.5cm,angle=-0]{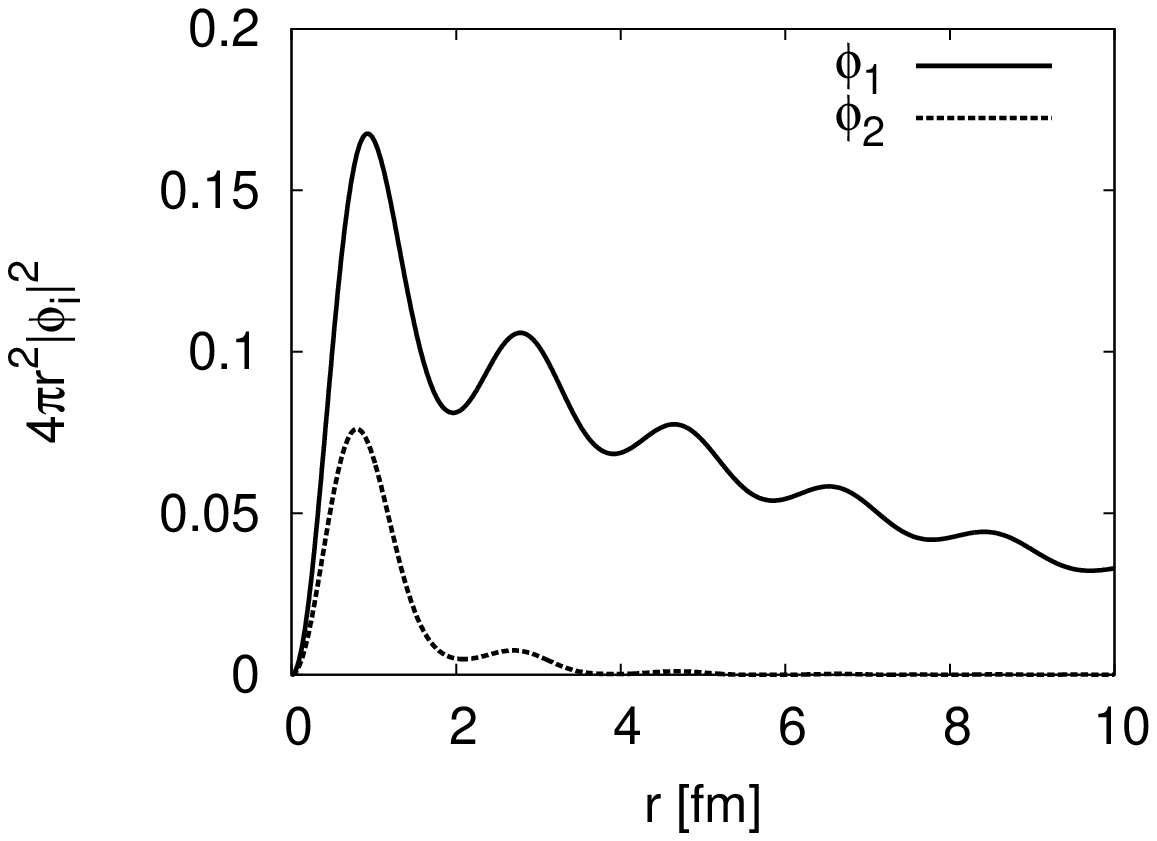} 
\end{tabular}
\caption{Neutral and charged Wave function components for a $\ddn$
  binding energy of 0.1 MeV. In the upper panel one can see the value
  of the wave functions at the origin for both channels, in the lower
  panel we plot the probability density for each channel.}
  \label{fig3}
\end{center}
\end{figure}

In Fig.~\ref{fig3} we show the two wave function components for the
problem solved in \cite{meutwox} with only two channels and the
binding energy for $\ddn$ of 0.1 MeV. The oscillations are caused by
the sharp cut-off $\Lambda$ introduced to regularize the Lippmann-Schwinger 
equation. 
Eqs.~(\ref{eq53a}) and (\ref{eq53b}) and the asymptotic behavior of
the wave functions given by Eqs.~(\ref{eq28'}) and (\ref{eq28''})
clearly show that, in the limit of zero binding energy of the $\ddn$
component, this wave function extends up to infinity while the $\ddc$
component is restricted in space because of the 8 MeV binding. The
probability to have the $\ddn$ component becomes much larger than that of the
$\ddc$ component (see \Eq{eq:ratio}) and we could think of the
$X(3872)$ as a $\ddn$ molecule. While technically correct from the
point of view of probabilities, this interpretation is misleading
concerning physical processes, like decays, because these require
Hamiltonians of short range, zero range ordinarily in effective field
theory, such that what matters in these processes is the wave function
at the origin. For instance, what will determine the $I=0$ character
of the wave function will be the $\psi_1(\vec{0})$ and
$\psi_2(\vec{0})$ magnitudes, not the probability integrated in all
coordinate space. Hence, Eqs.~(\ref{eq53a}) and (\ref{eq53b}) indicate
an isospin breaking, with respect to the $I=0$ combination, given by
the differences between $G_{11}^\alpha$ and $G_{22}^\alpha$ (see
Fig. \ref{fig2}).

Eq.~(\ref{rhoom}), related to the decay process depicted in
Fig.~\ref{fig1}, has now an intuitive representation to the light of
Eqs.~(\ref{eq53a}) and (\ref{eq53b}). The amplitude for
\be
X(3872)\rightarrow J/\psi \rho(\omega)
\ee 
is given by
\be
{\cal M}&=&g_1G_{11}^\alpha F_1+g_2G_{22}^\alpha F_2 \nn\\
&=&(2\pi)^{3/2}( \psi_1(\vec{0}\,)F_1+\psi_2(\vec{0}\,)F_2)\label{eq54}
\ee
where $F_1$ and $F_2$ are isospin factors for the vertices
$\ddn\rightarrow J/\psi \rho(\omega)$ and $\ddc\rightarrow J/\psi
\rho(\omega)$, $F_2/F_1=-1$ for the $\rho$ and $F_2/F_1=1$ for the
$\omega$. Certainly this should be the case for many channels and we
shall see the generalization in the next section.

Eq.~(\ref{eq54}) can alternatively be interpreted as
\be
{\cal M}&=&\int \dtres p \,\langle \pvec\,|\psi_1\rangle t_{\ddn\rightarrow J/\psi \rho(\omega)} \nn\\
&+&\int \dtres p \,\langle \pvec\,|\psi_2\rangle t_{\ddc\rightarrow J/\psi \rho(\omega)}\label{decayori}
\ee
assuming that the range of the $t$ amplitudes is very short compared to
the extension of the wave functions, essentially that the $t$ are
constant functions in momentum space, as one has in field theory
vertices stemming from a contact Lagrangian.

\section{Generalization to many channels}

We have now
\be
\langle \ppvec|V|\pvec\, \rangle&\equiv&v \Theta(\Lambda-\pvec\,) \Theta(\Lambda-\ppvec)
\ee
where $v$ is a $N\times N$ matrix with $N$ the number of channels.

The expressions that we have obtained can be generalized to many channels and we derive here some useful expressions.

\subsection{The couplings}

We can write for the $T$ matrix
\be
T&=&\frac{{\cal A}v}{\det(1-vG)} \label{eq56}
\ee
where ${\cal A}$ is defined as
\be
{\cal A}&=&\left[\,\det(1-vG)\right](1-vG)^{-1} \label{eq57}
\ee
This matrix is introduced to single out the source of the pole in all channels which is given by the condition
\be
\det(1-vG)&=&0\label{eq58}
\ee
We have now
\be 
g_ig_j&=&\lim_{E\rightarrow E_\alpha}(E-E_\alpha)T_{ij}\nn \\
&=& \left.\frac{({\cal
A}v)_{ij}}{\frac{d}{dE}\det(1-vG)}\right|_{E=E_\alpha} \label{eq59} \\
\frac{g_j}{g_i}&=&\left.\frac{({\cal A}v)_{ij}}{({\cal A}v)_{ii}}\right|_{E=E_\alpha}\label{eq60} 
\ee
We can see that $g_j/g_i$ is a ratio of two matrix elements of
matrices without singularities. This means that if $g_i\rightarrow 0$
as a consequence of having the binding in channel $i$ going to zero,
then all the couplings to the other channels coupled to channel $i$
will also go to zero. This is also obvious from Eq.~(\ref{eq59}) since
$g_j^2\rightarrow 0$ because  the denominator contains
$\frac{dG_{ii}^\alpha}{dE}$ and, one has
$\frac{dG_{ii}^\alpha}{dE}\rightarrow \infty$ for all cases.

\subsection{Wave functions}

Eqs.~(\ref{eq44a}) and (\ref{eq44b}) can be generalized as
\be
\langle \pvec\,|\psi_i\rangle &=&\Theta(\Lambda-p)\frac{1}{E-M_i-\frac{\vec{p}^{\,2}}{2\mu_i}} \nn \\
 &\times& \sum_j v_{ij}\int_{k<\Lambda} \dtres k \langle \kvec|\psi_j\rangle  \label{eq64}
\ee
which upon integration leads to 

\be
\int_{p<\Lambda} \dtres p\langle \pvec\,|\psi_i\rangle &=&G_{ii}\sum_j v_{ij}\nn\\
&\times&\int_{k<\Lambda} \dtres k \langle \kvec|\psi_j\rangle  \label{eq65}
\ee
which in the language of \Eq{eq53a} and \Eq{eq53b} reads
\be
\hat{\psi_i}&=&G_{ii} \sum_j v_{ij} \hat{\psi_j} \nn \\
\hat{\psi}&=&Gv\hat{\psi} \label{eq66}
\ee
where \Eq{eq66} is written in matrix form and it requires that
$\det(1-vG)=0$ for its solution, which is guaranteed for the bound
eigenstate. \Eq{eq66} can also be written as ($G^{-1}_\alpha = G^{-1}(E_\alpha)$)
\be
G^{-1}_\alpha\hat{\psi}&=&v\hat{\psi}\label{eq67}
\ee
which allows to rewrite the equation for the wave functions \Eq{eq64} as
\be
\langle \pvec\,|\psi\rangle &=&{\rm
  diag}\,\left(\frac{\Theta(\Lambda-p)}{E_\alpha-M_i-\frac{\vec{p}^{\,2}}{2\mu_i}}\right)G^{-1}_\alpha\hat{\psi} \label{eq68}
\ee
which gives the wave function in momentum space in terms of the wave
function in coordinate space at the origin. These equations are the
generalization of \Eq{eq48a} and \Eq{eq48b} together with \Eq{eq47}.

Let us now use the normalization condition
\be
\sum_i\langle \psi_i|\psi_i\rangle &=&\int \dtres p \sum_i|\langle \pvec\,|\psi_i\rangle |^2 \nn \\
&=&-\left.\sum_i \frac{dG_{ii}}{dE}\frac{1}{G_{ii}^2}\right|_{E=E_\alpha}\hspace{-0.5cm}\hat{\psi_i}^2=1 \label{eq70}
\\\nn
\ee

We can now take advantage of \Eq{eq68} to define the couplings in
terms of the $\hat{\psi_i}$. For this we use the version of \Eq{lseq2}
for the Lippmann Schwinger equation, recalling that close to the pole
of the eigenfunction of the Hamiltonian, $|\psi\rangle $, associated
to the energy $E_\alpha$, only this state contributes in the sum over
eigenstates of $H$, and we find
\be
T_{ij}&=&v_{ij}+\sum_{mn}v_{im}\int_{k<\Lambda}\dtres k\langle \kvec|\psi_m\rangle \nn\\
&\times&\frac{1}{E-E_\alpha}\int_{k^\prime<\Lambda}\dtres k^\prime\langle \kpvec|\psi_n\rangle v_{nj} \label{eq73}
\ee
which means that
\be
g_ig_j&=&\sum_{mn} v_{im}\hat{\psi}_m v_{nj}\hat{\psi}_n \nn\\
&=& \left.G^{-1}_{ii}\hat{\psi}_i G^{-1}_{jj}\hat{\psi}_j \right|_{E=E_\alpha} \label{eq74}
\ee
from where we conclude that 
\be
g_i&=&\hat{\psi_i} / G^{\alpha}_{ii} \nn \\
g_iG_{ii}^\alpha&=&\hat{\psi_i}\label{eq75}
\ee
as we found in \Eq{eq53a} and \Eq{eq53b} in the two channel
problem. This allows to reinterpret \Eq{eq70} in terms of the
couplings and we find
\be
\sum_i g_i^2 \left.\frac{dG_{ii}}{dE}\right|_{E=E_\alpha}=-1\label{eq76}
\ee
which is the generalization of \Eq{eqres}.

\Eq{eq76} is interesting because when one channel becomes loosely
bound then the loop derivative for this channel goes to infinity while
the other derivatives remain finite. In this limit we get, if channel
1 is loosely bound
\be 
\lim_{E_B\rightarrow 0}g_1^2\left.\frac{dG_{11}}{dE}\right|_{E=E_\alpha}&=&-1 \nn\\
g_1^2&=&-\left(\frac{dG_{11}}{dE}\right)^{-1}_{E=E_\alpha} \label{eq77}
\ee 
which is the same
result obtained for one channel in \Eq{eq25}. Thus, in this limit the
coupling of the loosely bound state goes to zero as the binding energy
goes to zero. On the other hand, \Eq{eq60} guarantees that all the
other couplings will also go to zero since the matrix ${\cal A}v$ is not
singular. This result was also found in \cite{jnieves} although
derived in a different way.

\section{Decay width of the $X(3872)$}
\label{sec:decay}

After these clarifications we would like to go back to the ratio of
the decay width of the $X(3872)$ to $J/\psi\rho$ and $J/\psi\omega$ of
\Eq{rhoom}. As discussed in \cite{meutwox}, the ratio of widths was
given by the square of \Eq{rhoom} times the factor to correct for the
phase space of $\rho$ decaying to two pions and the $\omega$ decaying
to three pions:
\begin{widetext}
\be \frac{{\cal B}(X\rightarrow J/\psi\pi\pi )}{{\cal B}(X\rightarrow
J/\psi\pi\pi\pi )}&=&\left(\frac{G^\alpha_{11}-G^\alpha
_{22}}{{G^\alpha_{11}+G^\alpha_{22}}}\right)^2 \frac{\int_0^{\infty }
q \mathcal{S}\left(s,m_{\rho },\Gamma _{\rho }\right) \Theta
\left(m_X-m_{J/\psi }-\sqrt{s}\right) \, ds}{\int_0^{\infty } q
\mathcal{S}\left(s,m_{\omega },\Gamma _{\omega }\right) \Theta
\left(m_X-m_{J/\psi }-\sqrt{s}\right) \, ds} \frac{{\cal
B}_\rho}{{\cal B}_\omega} \label{eq80} \ee
\end{widetext}
where ${{\cal B}_\rho}$ and ${{\cal B}_\rho}$ are the branching
fractions of $\rho$ decaying into two pions ($\sim$ 100 \%) and
$\omega$ decaying into three pions ($\sim$ 89 \%), $q$ is the center
of mass momentum of the outgoing meson pair in each channel and value
of $s$, and $\mathcal{S}\left(s,m,\Gamma\right)$ is the spectral
function of the mesons given by:
\be
\mathcal{S}\left(s,m,\Gamma\right)&=&-\frac{1}{\pi} {\rm Im}\left(\frac{1}{s-m^2+i \Gamma  m}\right)
\ee
In \cite{meutwox} it was found, using dimensional regularization for the loops,
\be
\frac{{\cal B}(X\rightarrow J/\psi\pi^+\pi^-\pi^0 )}{{\cal B}(X\rightarrow J/\psi\pi^+\pi^- )}&=&1.4 \label{eq81}
\ee
which is compatible with the experimental value $1.0\pm0.4$ from
\cite{belle2}.

Now let us assume that we take seriously that there is only one
channel, the $\ddn$. Then the ratio of the widths is
\be
R_{\rho/\omega}^{(\ddn)}&=&\left(\frac{\hat{\psi_1}t_{\ddn\rightarrow J/\psi\rho}}{\hat{\psi_1}t_{\ddn\rightarrow J/\psi\omega}}\right)^2=1\label{eqraml}
\ee
which is about 30 times bigger than the value obtained for this ratio
(0.032) in \cite{meutwox}\footnote{Note that 
\be
\left(\frac{G^\alpha_{11}-G^\alpha
  _{22}}{G^\alpha_{11}+G^\alpha_{22}}\right)^2 =
\left(\frac{1-\hat{\psi_2}/\hat{\psi_1}}
{1+\hat{\psi_2}/\hat{\psi_1}}\right)^2 
\ee
and thanks to Eq.~(\ref{eq:ratio-l}) the ratio of wave
functions at the origin depends little on the ultraviolet cutoff
$\Lambda$.}
. When we take into account the phase space
for the decay into $\rho$ and $\omega$ and the $\rho$ and $\omega$
branching ratios into two and three pions, with the ratio in
\Eq{eqraml} we find
\be
\frac{{\cal B}(X\rightarrow J/\psi\pi^+\pi^-\pi^0 )}{{\cal B}(X\rightarrow J/\psi\pi^+\pi^- )}&=&0.05\label{eq83}
\ee
which is about a factor 20 times smaller than experiment.

It is thus clear that the charged components of the wave function have
played an essential role bringing this branching ratio close to
experiment and this stresses that the wave functions at the origin for
each channel, and not the probabilities of finding the state in a
single channel alone, is what determines the isospin nature of the
state in coupled channels. Indeed, the $X(3872)$ wave
function would read
\begin{eqnarray}
\left. \langle \vec{r}\, | \psi \right>&=&
  \frac{\psi_1(\vec{r}\,)+
  \psi_2(\vec{r}\,)}{\sqrt{2}}\,\chi_{I=0}\nonumber\\
& + &  \frac{\psi_1(\vec{r}\,)- \psi_2(\vec{r}\,)}{\sqrt{2}}\,\chi_{I=1}
\end{eqnarray}
with $\chi_{I=0,1}$ scalar and vector isospin wave function
spinors. In the charge basis used in Section IV, the isospin wave functions are
\be
\chi_{I=0}=\frac{1}{\sqd}\left(\begin{array}{c} 1 \\ 1 \end{array}\right),&\hspace{.3cm} &
\chi_{I=1}=\frac{1}{\sqd}\left(\begin{array}{c} 1 \\ -1 \end{array}\right).
\ee
As mentioned in the introduction, the $X(3872)$ does not
have well defined isospin because the free Hamiltonian (including
masses) of the $\ddn$ and $\ddc$ system does not commute with
isospin. The mixing depends on the relative distance $\vec{r}$ between the
pseudoscalar and vector mesons. Thus
with transition operators of zero range, one
easily understands that the ratio of branching fractions of
Eq.~(\ref{ratiopipi}) is determined by the ratio
$\left[\left(1-\hat\psi_2/\hat\psi_1\right) /
\left(1+\hat\psi_2/\hat\psi_1\right)  \right]^2$, which in turn gives the
ratio of isospin 1 to isospin 0 probabilities at $\vec{r} =0 $. As we shall see in the next
section (see $R_{\rho/\omega}$ of Table \ref{tabcomp}), this ratio is of the order of 2\%.

\section{Results for other forms of the potential}
\label{sec:otherf}

One might think that the results obtained are specific of the type of
the potential chosen in \Eq{potte}, but the results are actually very
general.  To show that this is the case, we use other potentials. Let
us consider a separable potential where we substitute the sharp
cut off by a form factor
\be
\langle \ppvec|V|\pvec\, \rangle&\equiv&v f(\pvec\,) f(\ppvec)
\ee
where $v$ is a $N\times N$ matrix with $N$ the number of channels.
The results of Sect.~V follow nearly identically substituting the
$\Theta(\Lambda-p)$ by $f(\pvec\,)$. \Eq{eq56}-\Eq{eq60} are the same, 
but $G$ is now given by
\be
G_{ii}&=&\int \dtres p f^2(\pvec\,) \frac{1}{E-M_i-\frac{\pvec^{\,2}}{2\mu_i}}
\ee
and the wave functions are now given by
\be
\langle\pvec\,|\psi_i\rangle&=&f(\pvec\,)
\frac{1}{E_\alpha-M_i-\frac{\pvec^{\,2}}{2\mu_i}}\sum_j v_{ij}\nn\\
&\times&\int \dtres k f(\kvec)\langle\kvec|\psi_j\rangle \label{wfnp}
\ee
\Eq{eq65}-\Eq{eq67} follow, but now
\be
{\hat \psi}_i&=&\int \dtres k f(\kvec)\langle\kvec|\psi_i\rangle \label{psihatn}
\ee
which allows to write \Eq{wfnp} as
\be
\langle\pvec\,|\psi\rangle&=&{\rm
  diag}\left(\frac{f(\pvec\,)}{E_\alpha-M_i-\frac{\pvec^{\, 2}}{2\mu_i}}\right)G^{-1}_\alpha{\hat \psi}\label{wfdiag}
\ee
and again we find \Eq{eq75}
\be
g_i&=&{\hat \psi}_i/G_{ii}^\alpha \nn\\
g_i G_{ii}^\alpha&=&{\hat\psi}_i \label{gipsi}
\ee
and \Eq{eq76}-\Eq{eq77} also follow.

Everything is identical as before, but now ${\hat \psi}$ is not, up to
a factor $(2\pi)^\frac32$ , the
wave function at the origin (it would be if we removed $f(\pvec\,)$
from \Eq{psihatn}).  To see the meaning of ${\hat\psi}$ we write
$f(\pvec\,)$ in terms of its Fourier Transform
\be
f(\pvec\,)&=&\frac{1}{(2\pi)^{3/2}}\int \dtres x {\hat f}(\xvec)e^{i\pvec.\xvec}
\ee
and the wave function of \Eq{wfdiag} also in terms of its Fourier
Transform
\be
\psi_i(\pvec\,)&=&\frac{1}{(2\pi)^{3/2}}\int \dtres x e^{-i\pvec.\xvec}\psi_i(\xvec)
\ee
Then upon integrating explicitly over $\kvec$ in \Eq{psihatn} we find
\be
{\hat\psi}_i&=&\int \dtres x \, \psi_i(\xvec) {\hat f}(\xvec)
\ee
We performed explicit calculations using a gaussian form for $f(\pvec)$
\be
f(\pvec\,)&=&e^{-\frac{1}{2}\pvec^{\,2}/\Lambda^2} \nn\\
{\hat f}(\xvec)&=&\Lambda^3e^{-\frac{1}{2}\xvec^2\Lambda^2}
\ee
and a Lorentz form 
\be 
f(\pvec)&=& \frac{\Lambda^2}{\Lambda^2+\pvec^{\,2}} \nn\\
{\hat f}(\xvec)&=&\sqrt{\frac{\pi}{2}}\Lambda^2\,
\frac{e^{-|\xvec|\Lambda}}{|\xvec|} 
\ee
We can see that ${\hat f}(\xvec)$ has a range of $1/\Lambda\sim
0.2-0.3$ fm, a range much smaller than the extension of the wave
function. Thus ${\hat \psi}_i$ gives the average of the wave function
in the vicinity of the origin, while in the case of the sharp cut off
one finds exactly the wave function at the origin.

One sees again that this average value of the wave function at the origin is 
what governs the decay process of the resonance. Indeed, if we go to
\Eq{decayori} to get the amplitudes for the decay mechanism of Fig. \ref{fig1}
we would find:
\be
{\cal M}&=&\int \dtres p \,\langle \pvec\,|\psi_1\rangle f(\pvec\,)\,t_{\ddn\rightarrow J/\psi \rho(\omega)} \nn\\
&+&\int \dtres p \,\langle \pvec\,|\psi_2\rangle f(\pvec\,)\,t_{\ddc\rightarrow J/\psi \rho(\omega)}\label{eqMn}
\ee
where we have taken for the $\ddi\rightarrow J/\psi\rho(\omega)$ transition
amplitude, for consistency, $f(\pvec)t_{\ddi\rightarrow J/\psi \rho(\omega)}$ with $t$ a constant.
\Eq{eqMn} reads
\be
{\cal M}&=&g_1G_{11}^\alpha t_{\ddn\rightarrow J/\psi \rho(\omega)}\nn
\\
&+& g_2G_{22}^\alpha t_{\ddc\rightarrow J/\psi \rho(\omega)}
\ee
as we obtained before, and according to \Eq{gipsi} now $g_i
G_{ii}={\hat \psi}_i$.

It is interesting to compare the results with the form factor with
those we had before. We proceed as follows, we take the same strength
for the potential as before and determine $\Lambda$ to get the neutral
channel bound $\ddn$ at the same energy.  In Table \ref{tabcomp} we show
the values of $g$, $\psihat$ and $\Lambda$ for the three approaches: the
sharp cut off, a gaussian form factor and a lorentzian form factor.
We see that the differences for $g$ and $\psihat_i$, the relevant magnitudes
in the $X(3872)$ decay, are small. Note that the value of the $R_{\rho/\omega}$
of \cite{meutwox} was obtained using dimensional regularization and zero binding energy
for the $\ddn$ channel. If the calculation is made in the same scheme but
with $E_B$=0.1 MeV one obtains $R_{\rho/\omega}$=0.025, which differs from the values
on Table \ref{tabcomp}. The bulk of this difference can be attributed to differences
between using hard cut off in momentum space or dimensional regularization.
Indeed, using the relativistic treatment of \cite{meutwox}, but using a hard cut off,
provides a value for $R_{\rho/\omega}$=0.020 at $E_B=$0.1 MeV.

\begin{figure}
\begin{center}
\begin{tabular}{c}
\includegraphics[width=7.5cm,angle=-0]{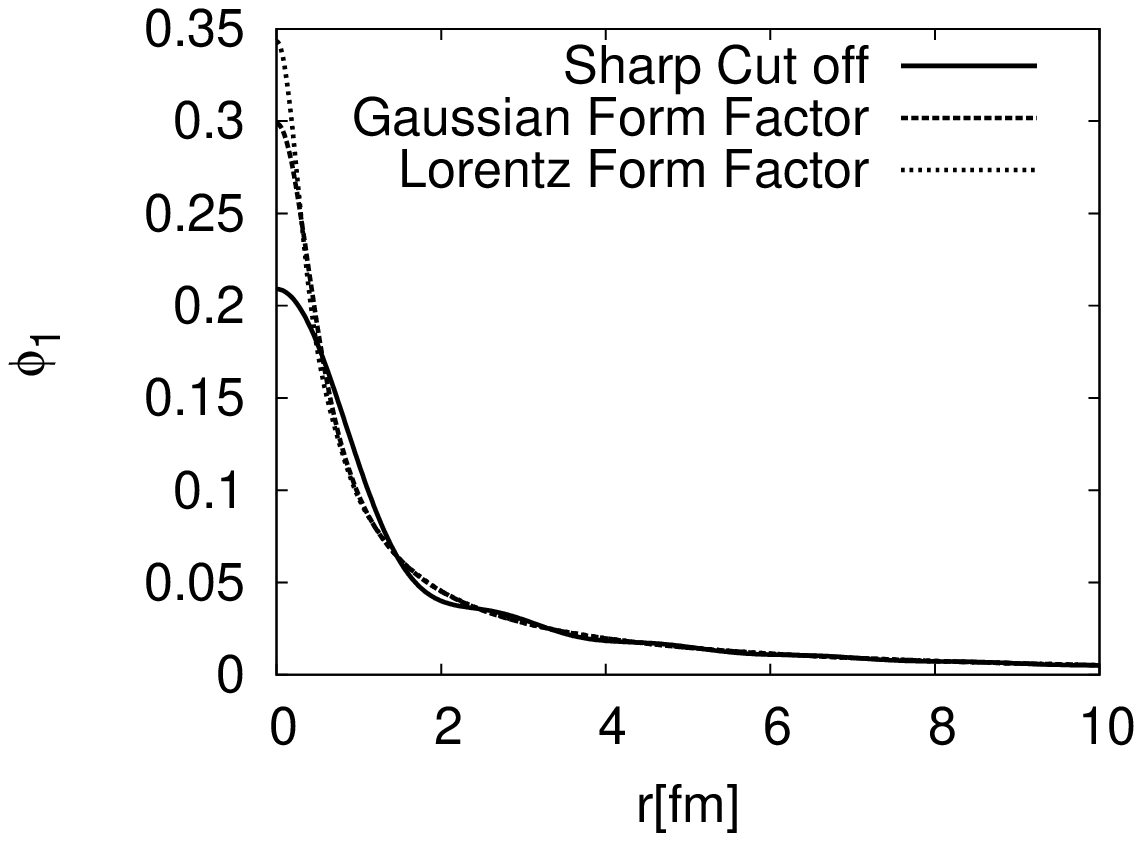} \\
\includegraphics[width=7.5cm,angle=-0]{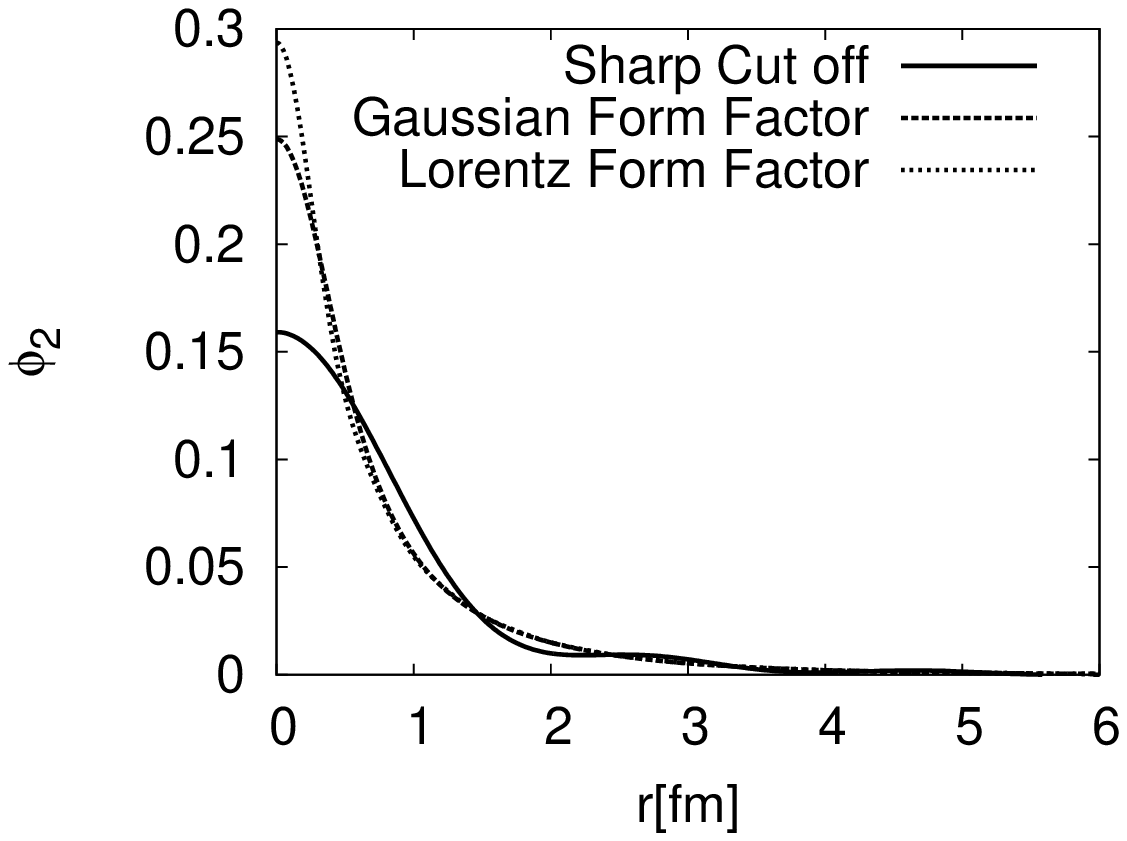}
\end{tabular}
\caption{Wave functions for different form factors in the potential.}
  \label{wavefunccomp}
\end{center}
\end{figure}

We also plot in Fig. \ref{wavefunccomp} the wave functions obtained with the three
approaches and, as we see, they hardly differ above $0.5-0.6~ {\rm fm}$.
The differences at the origin cancel in $\psihat_i$, where a smearing around the origin is involved,
and particularly in the ratios $\hat \psi_1 / \hat \psi_2$.

The robustness of the results against changes in the form factor suggests
that a large cut-off approximation should work fairly well. Indeed, in this limit
all cut-off procedures merge into a single one, and taking into account
that $G_{ii}(E_\alpha)-G_{ii}(M_i)$ is finite we get
\be 
\hspace{-0.3cm}G_{ii}(E_\alpha) &=& - 8\pi \mu_i \bigg[ \int_0^\infty dp [f(\pvec\,)]^2 -
  \frac{\pi}2 \gamma_i \nn\\
  &+& {\cal O} (\Lambda^{-1}) \bigg] \label{eq:G-large}
\ee
which shows that what matters is the integrated strength of
$[f(\pvec)]^2$ and corresponds to using a common subtraction constant
with a different cut-off interpretation, 
\be
\int_0^\infty dp [f(\pvec\,)]^2 &=& \Lambda_{\rm Sharp} = \frac{\sqrt{\pi}}2 \Lambda_{\rm Gauss} = \frac{\pi}4 \Lambda_{\rm Lorentz}  \nonumber \\ 
\ee 
This identification provides a simple rule relating the different
cut-offs $\Lambda_{\rm Sharp}$, $ \Lambda_{\rm Gauss}$ and
$\Lambda_{\rm Lorentz}$ which works very well as can be checked from
Table~\ref{tabcomp}.
Keeping the leading terms in
Eq.~(\ref{eq:G-large}) and taking $\mu_1= \mu_2 = \bar \mu$ we obtain
the following remarkably simple analytical results by using the bound state
condition, \Eq{eq46},
\be
\int_0^\infty dp [f(\pvec)]^2 &=& \frac{\pi}{4} (\gamma_1 + \gamma_2)
+ \frac{2 \pi^2 f_D^2}{\bar \mu} \\
R_{\rho/\omega}  &=& \frac{(\gamma_1 - \gamma_2)^2 \bar \mu^2}{64 f_D^4\pi^2}
\ee
which yields
\be
\int_0^\infty dp [f(\pvec)]^2= 665 {\rm MeV} \, , \quad R_{\rho/\omega}=0.025
\ee
Next we want to connect the present results with those obtained in \cite{meuax,meutwox}
in the relativistic approach. Actually, by matching the relativistic one-loop integral calculated within
dimensional regularization with scale $\nu$ (called $\mu$ in ~\cite{meuax}) with the
non-relativistic propagator in the heavy meson limit $m_D, m_{D^*} \gg
\gamma_1,\gamma_2$ (see Eqs.~(\ref{eq:jnb}), (\ref{eq:jn1})
and (\ref{eq:G-large})), we get\footnote{We are using averaged values of masses for $D$ and $D^*$ in this formula.
This is appropriate since no differences of masses appear in the formula.
This allows the matching between relativistic and non-relativistic expressions.}
\be
\int_0^\infty dp [f(\pvec)]^2 &=&    
-\frac14 \Big[ (m_D+m_{D^*}) \alpha_{H} \nonumber \\  
&+& m_D \log(\frac{m_D^2}{\nu^2})+m_{D^*} \log(\frac{m_{D^*}^2}{\nu^2})
\Big] 
\ee 
where $\alpha_{H}$ is a dimensionless subtraction constant which
depends on the scale $\nu$. For $\nu =1.5$ GeV the
subtraction constant used in Ref.~\cite{meutwox} was $\alpha_{H} = -1.185$ yielding $\int_0^\infty dp
[f(\pvec)]^2 = 651 \,{\rm MeV}$ in fairly good agreement with
Table~\ref{tabcomp} ($\Lambda_{\rm Sharp}$).  Neglecting the finite cut-off corrections as in
Eq.~(\ref{eq:G-large}) one obtains now $R_{\rho /\omega}=0.026$.

\begin{table}
\begin{center}
\caption{Comparative results for different potentials for a $\ddn$
  binding energy of 0.1 MeV.} \label{tabcomp}
\vspace{0.5cm}
\begin{tabular}{c|c|c|c|c|c|c|c}
Form & $\Lambda$ & $g^{\textrm{FT}}$ & $\psi_1(\vec{0}\,)/\psi_2(\vec{0}\,)$ & $\psihat_1/\psihat_2$ & $\psihat_1$& $\psihat_2$& $R_{\rho/\omega}$ \\
Factor & [MeV] & [MeV] & & & & &\\
\hline
\hline 
Sharp & 653 & 3202 & 1.31 & 1.31 &3.29&2.50& 0.018 \\
\hline
Gauss & 731 & 3238 & 1.20 & 1.29 &3.30&2.56& 0.016 \\
\hline
Lorentz & 834 & 3254 & 1.17 & 1.28 &3.31&2.58& 0.015
\end{tabular}
\end{center}
\end{table}

\section{Conclusions}

With a view to the structure of the $X(3872)$ as a possible coupled
channel bound state of mostly the $\ddn$ and $\ddc$ and other minor
channels, we have studied the meaning of the couplings, which one
determines from the residues of the scattering matrix at the poles, in
terms of wave functions for the different channels. We have done the
study in one channel, then in two channels suited to the $X(3872)$
resonance and then we have generalized the results to many
channels. Interesting relationships are obtained which shed light on
the field theoretical approaches to reactions from the perspective of
wave functions. Essentially we find that the couplings are
proportional to the value of the wave function at the origin in
coordinate space or the averaged value within the range of the interaction.
They are not sensitive to the wave function at long distances which is governed by the binding energy,
and we also find that this averaged value of the wave function at the origin,
$\psihat=gG$, is the only information
that is needed when dealing with short range processes, like those
provided in terms of contact Lagrangians in field theory.
We also found that the values $\psihat_i$ were very stable against
assumed shapes of the potential once the binding energy is fixed fulfilling
the quantization condition $det(1-VG)=0$.

We also find that, when one channel becomes loosely bound, the couplings
to all coupled channels go to zero. Even if in terms of probabilities
the loosely bound channel, whose wave function extends up to
infinity, has the largest probability, what matters in the reactions
is the averaged values of the wave functions at the origin that
determine the dynamics of the processes and the underlying
symmetries like isospin. The isospin violation in particular is
tied to the ratio of wave functions around the origin $\psihat_1/\psihat_2$ (for short range processes),
which goes to a finite limit when the binding of the $\psi_1$ component goes to zero.

When coming to the $X(3872)$ case, which can correspond to the $\ddn$
channel very loosely bound and the $\ddc$ bound by about 8 MeV, we
find that the wave functions at the origin for the two channels are
similar, suggesting that one has a state with $I=0$, with
small isospin breaking, even if the probability to find the $\ddn$
component in the full space is much larger than for the $\ddc$
component. A precise measure of the isospin admixture is given by
the ratio $\psihat_1/\psihat_2$, which is very stable and has a value around 1.3,
the value of 1 corresponding to a pure $I=0$ state where the decay
$X\rightarrow J/\psi\rho$ would be forbidden.

The consideration of the charged $\ddc$ component to describe
physical processes is so important that if it is neglected one finds a
ratio of $\frac{{\cal B}(X\rightarrow J/\psi\pi^+\pi^- )}{{\cal
B}(X\rightarrow J/\psi\pi^+\pi^-\pi^0 )}$ twenty times bigger than
experiment.

The work done has also an academical component. Some useful
expressions, as well as exact analytical solutions for the wave
functions in coupled channels have been given. The work also shows a
different perspective on the on-shell approach to the scattering
matrix based on the N/D method used in all modern works of chiral
dynamics in coupled channels, by means of which the coupled Bethe
Salpeter integral equations become algebraic ones. The suitable choice
of the potential in momentum space that we made gives rise to the same
equations as in the field theoretical on-shell approach. The
analytical expressions found can be very useful to give alternative
interpretations of results found in the chiral unitary approach, or in
general in unitary coupled channels methods in many physical
processes.


\section*{Acknowledgments}

This work is partly supported by DGI and FEDER funds, under contract
FIS2006-03438, FIS2008-01143/FIS and PIE-CSIC 200850I238 and the Junta
de Andalucia grant no. FQM225-05. We acknowledge the support of the
European Community-Research Infrastructure Integrating Activity "Study
of Strongly Interacting Matter" (acronym HadronPhysics2, Grant
Agreement n. 227431) under the Seventh Framework Programme of EU.
Work supported in part by DFG (SFB/TR 16, "Subnuclear Structure of
Matter").




\begin{thebibliography}{99}


\bibitem{belle}
  S.~K.~Choi {\it et al.}  [Belle Collaboration],
  Phys.\ Rev.\ Lett.\  {\bf 91}, 262001 (2003)
  [arXiv:hep-ex/0309032].

\bibitem{cdf}
  D.~E.~Acosta {\it et al.}  [CDF II Collaboration],
  Phys.\ Rev.\ Lett.\  {\bf 93}, 072001 (2004)
  [arXiv:hep-ex/0312021].

\bibitem{d0}
  V.~M.~Abazov {\it et al.}  [D0 Collaboration],
  Phys.\ Rev.\ Lett.\  {\bf 93}, 162002 (2004)
  [arXiv:hep-ex/0405004].

\bibitem{babar}
  B.~Aubert {\it et al.}  [BABAR Collaboration],
  Phys.\ Rev.\  D {\bf 71}, 071103 (2005)
  [arXiv:hep-ex/0406022].
  
\bibitem{badhonef}
 G.~Bali {\it et al.},
 arXiv:0910.3165 [hep-ph].
  
\bibitem{qq1}
  E.~S.~Swanson,
  Phys.\ Rept.\  {\bf 429}, 243 (2006)
  [arXiv:hep-ph/0601110].

\bibitem{qq2}
  G.~Bauer,
  Int.\ J.\ Mod.\ Phys.\  A {\bf 21}, 959 (2006)
  [arXiv:hep-ex/0505083].

\bibitem{qq3}
  M.~B.~Voloshin,
  Prog.\ Part.\ Nucl.\ Phys.\  {\bf 61}, 455 (2008)
  [arXiv:0711.4556 [hep-ph]].
  
\bibitem{tornq}
  N.~A.~Tornqvist,
  Phys.\ Lett.\  B {\bf 590}, 209 (2004)
  [arXiv:hep-ph/0402237].

\bibitem{suzuki}
  M.~Suzuki,
  Phys.\ Rev.\  D {\bf 72}, 114013 (2005)
  [arXiv:hep-ph/0508258].

\bibitem{closeypage}
  F.~E.~Close and P.~R.~Page,
  Phys.\ Lett.\  B {\bf 578}, 119 (2004)
  [arXiv:hep-ph/0309253].

\bibitem{meuax}
  D.~Gamermann and E.~Oset,
  Eur.\ Phys.\ J.\  A {\bf 33}, 119 (2007)
  [arXiv:0704.2314 [hep-ph]].

\bibitem{meutwox}
  D.~Gamermann and E.~Oset,
  Phys.\ Rev.\  D {\bf 80}, 014003 (2009)
  [arXiv:0905.0402 [hep-ph]].

\bibitem{mol1}
  Y.~R.~Liu, X.~Liu, W.~Z.~Deng and S.~L.~Zhu,
  Eur.\ Phys.\ J.\  C {\bf 56}, 63 (2008)
  [arXiv:0801.3540 [hep-ph]].
  
\bibitem{mol2}
  X.~Liu, Y.~R.~Liu, W.~Z.~Deng and S.~L.~Zhu,
  Phys.\ Rev.\  D {\bf 77}, 034003 (2008)
  [arXiv:0711.0494 [hep-ph]].
  
\bibitem{mol3}
  Y.~b.~Dong, A.~Faessler, T.~Gutsche and V.~E.~Lyubovitskij,
  Phys.\ Rev.\  D {\bf 77}, 094013 (2008)
  [arXiv:0802.3610 [hep-ph]].
  
\bibitem{mol4}
  E.~S.~Swanson,
  Phys.\ Lett.\  B {\bf 588}, 189 (2004)
  [arXiv:hep-ph/0311229].

\bibitem{mol5}
  M.~B.~Voloshin,
  Phys.\ Lett.\  B {\bf 604}, 69 (2004)
  [arXiv:hep-ph/0408321].

\bibitem{mol6}
  E.~Braaten and M.~Kusunoki,
  Phys.\ Rev.\  D {\bf 72}, 054022 (2005)
  [arXiv:hep-ph/0507163].
  
\bibitem{mol7}
  X.~Liu, Y.~R.~Liu and W.~Z.~Deng,
  arXiv:0802.3157 [hep-ph].

\bibitem{mol8}
  Y.~Dong, A.~Faessler, T.~Gutsche, S.~Kovalenko and V.~E.~Lyubovitskij,
  arXiv:0903.5416 [].
  
\bibitem{mol9} E. Braaten, talk at the International Workshop on 
 Effective Field Theories: from the Pion to the Upsilon.
 http://ific.uv.es/eft09/
 
\bibitem{mol10}
  E.~Braaten and M.~Lu,
  Phys.\ Rev.\  D {\bf 76}, 094028 (2007)
  [arXiv:0709.2697 [hep-ph]].

\bibitem{mol11}
  M.~B.~Voloshin,
  Phys.\ Lett.\  B {\bf 579}, 316 (2004)
  [arXiv:hep-ph/0309307].

\bibitem{mol12}
  E.~Braaten and M.~Lu,
  Phys.\ Rev.\  D {\bf 77}, 014029 (2008)
  [arXiv:0710.5482 [hep-ph]].

\bibitem{Matheus:2009vq}
  R.~D.~Matheus, F.~S.~Navarra, M.~Nielsen and C.~M.~Zanetti,
  Phys.\ Rev.\  D {\bf 80}, 056002 (2009)
  [arXiv:0907.2683 [hep-ph]].

\bibitem{Lee:2009hy}
  I.~W.~Lee, A.~Faessler, T.~Gutsche and V.~E.~Lyubovitskij,
  arXiv:0910.1009 [Unknown].

\bibitem{hanhart}
  C.~Hanhart, Yu.~S.~Kalashnikova, A.~E.~Kudryavtsev and A.~V.~Nefediev,
  Phys.\ Rev.\  D {\bf 76}, 034007 (2007)
  [arXiv:0704.0605 [hep-ph]].
  
\bibitem{thomas}
  C.~E.~Thomas and F.~E.~Close,
  Phys.\ Rev.\  D {\bf 78}, 034007 (2008)
  [arXiv:0805.3653 [hep-ph]].
  
\bibitem{noverd}
  J.~A.~Oller and E.~Oset,
  Phys.\ Rev.\  D {\bf 60}, 074023 (1999)
  [arXiv:hep-ph/9809337].

\bibitem{meissner}
  J.~A.~Oller and U.~G.~Meissner,
  Phys.\ Lett.\  B {\bf 500}, 263 (2001)
  [arXiv:hep-ph/0011146].

\bibitem{Nieves:1999bx}
  J.~Nieves and E.~Ruiz Arriola,
  Nucl.\ Phys.\  A {\bf 679}, 57 (2000)
  [arXiv:hep-ph/9907469].

  
\bibitem{cdf2}
  A.~Abulencia {\it et al.}  [CDF Collaboration],
  Phys.\ Rev.\ Lett.\  {\bf 96}, 102002 (2006)
  [arXiv:hep-ex/0512074].

\bibitem{belle2}
  K.~Abe {\it et al.},
  arXiv:hep-ex/0505037.

\bibitem{osetreview}
  J.~A.~Oller, E.~Oset and A.~Ramos,
  Prog.\ Part.\ Nucl.\ Phys.\  {\bf 45}, 157 (2000)
  [arXiv:hep-ph/0002193]. 
  
\bibitem{hidden1}
  M.~Bando, T.~Kugo, S.~Uehara, K.~Yamawaki and T.~Yanagida,
  Phys.\ Rev.\ Lett.\  {\bf 54}, 1215 (1985).

\bibitem{hidden2}
  M.~Bando, T.~Kugo and K.~Yamawaki,
  Phys.\ Rept.\  {\bf 164}, 217 (1988).

\bibitem{hidden3}
  M.~Harada and K.~Yamawaki,
  Phys.\ Rept.\  {\bf 381}, 1 (2003)
  [arXiv:hep-ph/0302103].


\bibitem{weinberg}
  S.~Weinberg,
  Phys.\ Rev.\  {\bf 130}, 776 (1963).
  
\bibitem{weinberg2}
  S.~Weinberg,
  Phys.\ Rev.\  {\bf 137}, B672 (1965).



\bibitem{baru}
  V.~Baru, J.~Haidenbauer, C.~Hanhart, Yu.~Kalashnikova and A.~E.~Kudryavtsev,
  Phys.\ Lett.\  B {\bf 586}, 53 (2004)
  [arXiv:hep-ph/0308129].
  
  
\bibitem{jnieves}
  H.~Toki, C.~Garcia-Recio and J.~Nieves,
  Phys.\ Rev.\  D {\bf 77}, 034001 (2008)
  [arXiv:0711.3536 [hep-ph]].



\end{thebibliography}
\end{document}